\documentstyle[11pt,epsf]{article}
\newcommand{\be}{\begin{equation}}
\newcommand{\ee}{\end{equation}}
\newcommand{\bea}{\begin{eqnarray}}
\newcommand{\eea}{\end{eqnarray}}

\begin{document}
\thispagestyle{empty}
\setcounter{page}{0}

\begin{flushright}
UCY - PHY - 96/7 \\
ADP - 97 - 004/T243 \\
PSI - PR - 97 - 04
\end{flushright}

\vspace{1.5cm}
 
\renewcommand{\thefootnote}{\fnsymbol{footnote}}
\begin{center}
{\large \bf Variational Field Theoretic Approach to Relativistic Scattering  }
 
\vspace{0.5cm}

C.~Alexandrou $^{1 \>, \> 2}$, R.~Rosenfelder $^2$
and A.~W.~Schreiber $^{2 \>, \> 3, \> 4}$

\vspace{1cm}
$^1$ Department of Natural Sciences, University of Cyprus, CY-1678 Nicosia, 
Cyprus

$^2$ Paul Scherrer Institute, CH-5232 Villigen PSI, Switzerland

$^3$ TRIUMF, 4004 Wesbrook Mall, Vancouver, B.C.,
Canada V6T 2A3

$^4$ Department of Physics and Mathematical Physics and
Research Centre for the Subatomic Structure of Matter, University of Adelaide, 
Adelaide, S. A. 5005, Australia  

\end{center} 

\renewcommand{\thefootnote}{\arabic{footnote}}
\setcounter{footnote}{0}
 
\vspace{3cm}
\begin{abstract}
\noindent
Nonperturbative polaron variational methods are applied, within the
so-called particle or worldline representation of relativistic field
theory, to study scattering in the context of the scalar Wick - Cutkosky
model. Important features of the variational calculation are that it
is a controlled approximation scheme valid for arbitrary coupling
strengths, the Green functions have all the cuts and poles expected
for the exact result at any order in perturbation theory and that the
variational parameters are simultaneously sensitive to the infrared as
well as the ultraviolet behaviour of the theory.  
We generalize the previously used quadratic trial action
by allowing more freedom for off-shell propagation
without a change in the on-shell variational equations and evaluate
the scattering amplitude at first order in the variational scheme. 
Particular attention
is paid to the $s$-channel scattering near threshold because here
non-perturbative effects can be large.  
We check the unitarity of a our numerical calculation and find
it greatly improved compared to perturbation theory and to the zeroth order 
variational results.
\end{abstract}                     
 
\newpage
 
\setcounter{equation}{0}
 
\section{Introduction}

Variational techniques are widely used in the study of quantum
mechanical systems. Most of these applications are based on the
Rayleigh - Ritz variational principle for the energy of the system and
physical intuition plays a crucial role in the construction of the
variational wave function which is used. Applications of similar
techniques to relativistic quantum field theories are, in contrast,
rather limited. Intuitive arguments are hampered by having to deal
with infinitely many degrees of freedom and the problem is further
complicated by the need for renormalization. However, a well known
example of a non-relativistic field theory where the variational
approach has been highly successful is the polaron problem in
condensed matter physics~\cite{Fey1,FeHi}.  Here it yields results
within a few \% compared to exact Monte Carlo results~\cite{AlRo}.
The main idea on which this success is based is the exact
integration of the phonon degrees of freedom, thus obtaining an
effective one-body system where an intuitive construction of a trial
action is possible.  The same approach has been applied to the Walecka
model to study the one-~\cite{Pol1} as well as many-~\cite{Pol2}
nucleon system.

The generalization of this method to relativistic theories is possible
and forms the basis of the work described in this paper.  It should be
stressed that this is a variational method which is not related to the
usual variational approach in quantum field theory, the latter using trial
wave functionals of the field in order to obtain an approximation to
the effective action (see~\cite{BaGh} and references therein).  Exact
elimination of the light degrees of freedom (analogous to the phonons
in the polaron problem) remains a central idea in this application.
Furthermore, it is useful to employ the {\it particle} or {\it
worldline} representation [2, 7 - 10] for the
remaining heavy degrees of freedom of the theory.  This formulation
employs particle trajectories, parameterized in terms of a proper
time, rather than local fields.  Through its use the relativistic
theory bears a startling similarity to the non-relativistic polaron
problem and so it is not unreasonable to expect that some of the
success enjoyed by the variational method so far will be repeated.

In a recent series of papers [11-14]
(henceforth referred to as I -- IV) we developed the variational
treatment of a simplified field theory of two scalar
interacting particles.  The Lagrangian of this theory is of the
Wick-Cutkosky type~\cite{Wick,Cut} with a finite meson mass. As such,
it has the same type of interaction as Lagrangians relevant to
physical situations, for example any meson-nucleon theory or, in the
limit of zero-mass, Quantum Electrodynamics (QED).  On the other hand, 
it avoids the complications of a gauge theory as well as spin, chiral symmetry,
etc. so it is fair to say that its primary purpose is to serve as a
`playground' on which to develop the necessary techniques of dealing
with the strong coupling problem.  Indeed, it is for this reason that,
even after more than 40 years since Wick and Cutkosky first used the
model to look for the relativistic bound state in the ladder
approximation, it retains its
popularity [17 - 23]. The generalization of
the variational method to a realistic theory, namely QED, will be
presented in a forthcoming publication \cite{QED} (for a preliminary 
account see \cite{Dubna}).

The Wick - Cutkosky Lagrangian in Euclidean space is given by
\be
{\cal L} = \frac{1}{2}(\partial_{\mu} \Phi)^2+\frac{1}{2}M_0^2\Phi^2
+ \frac{1}{2}(\partial_{\mu} \phi)^2 +\frac{1}{2}m^2\phi^2-g\Phi^2\phi
\ee
where $M_0$ is the bare mass of the heavy particle $\Phi$, $m$ is the
mass of the light particle  $\phi$ and $g$ is the coupling constant.
As in (I) -- (IV), we work in the quenched approximation of the theory and
shall concentrate on those Green functions which involve one heavy particle
interacting with an arbitrary number of light particles.  The generating
functional for these Green functions is obtained after integration over the
light degrees of freedom and is given by
 (see (I) and (II) for details) 
\be
 Z[j,x] \propto \int_0^{\infty} d\beta \exp\left(-\frac{\beta}{2} M_0^2\right)
\int_{x(0)=0}^{x(\beta)=x} {\cal D}x(t) \exp \left \{ \> -S_{\rm
eff}[x(t)]-S_2[x(t),j]\> \right \}  \quad.
\ee
Here the effective action   
\be
S_{\rm eff} \> = \> \int_0^{\beta} dt  \> \frac{1}{2}\dot{x}^2(t)
-\frac{g^2}{2}\int_0^{\beta} \>\> dt_1 \int_0^{\beta} \>\> dt_2
\int \frac{d^4q}{(2\pi)^4} \>\frac{1}{q^2+m^2}\>
e^{iq \cdot (x(t_1)-x(t_2))}
\label{eq: effective action}
\ee
is formulated in terms of the heavy particle trajectories parameterized by 
the proper time $t$.
The meson source term 
\be
S_2[x(t),j] \> = \> -g\int d^4y \> j(y)\>\> \int_0^{\beta} \> dt \int
\frac{d^4q}{(2\pi)^4} \>\>\frac{e^{iq \cdot (y-x(t))}}{q^2+m^2}
\ee
produces $n$ external mesons upon differentiation of  $Z[j,x]$ with respect 
to the meson source $j(y)$ and setting $j(y) = 0$.

The variational method due to Feynman is based on the construction of a 
suitable 
trial action $S_t$ containing variational parameters
and on Jensen's inequality (or stationarity for complex actions)
\be
<e^{-\Delta S}>_{S_t} \stackrel{\rm stat}{\simeq} e^{-<\Delta S>_{S_t}}\;\;\;,
\label{eq: jensen}
\ee 
where
\be
 \Delta S=S_{\rm eff}-S_t
\ee
and the averaging is done with the weight function $e^{-S_t}$. This
averaging must be done exactly and therefore the  trial  action can depend
at most quadratically on the function $x(t)$.

Several remarks are in order here.  First of all, one may wonder in
what way a quadratic trial action could in any way be a reasonable
approximation to the effective action in 
Eq.~(\ref{eq: effective action}) which, after all, is exponential in $x(t)$.  
The important
feature of $S_{\rm eff}$ (which, by construction, is shared by the
trial action $S_t$) is that it is non-local in the proper time $t$:
For an `average' path $x(t)$ the separation $x(t_1) - x(t_2)$ tends
to be correlated with the proper time difference $t_1 - t_2$ and
so an unrestrained increase due to a quadratic term 
$ \> \left [ \>x(t_1) - x(t_2) \> \right ]^2 \> $ in the trial action may 
be compensated for by a
`retardation function' in $t_1 - t_2$.  Secondly, it is
important to note that the approximation in Eq.~(\ref{eq: jensen}) is a
correctable one: if one so desires one can calculate arbitrarily
precise corrections to the variational result by adding appropriate
powers of $\Delta S$ on the right hand side of this equation.  

Another important feature of Eq.~(\ref{eq: jensen}) is that it is easy
to ensure that it reproduces the one loop perturbative result for {\it
any} trial action whatsoever (see Ref. (I)).  It goes without saying,
however, that the more of the physics the trial action contains, the
closer the numerical results of the `leading order' result in
Eq.~(\ref{eq: jensen}) should be to the true situation. In particular,
in the present paper we shall examine meson-`nucleon' scattering near
threshold (for simplicity, we refer to the heavier particle as the
`nucleon' and the lighter as the meson).  One expects large distance (time)
effects to be important at threshold since the particles move very slowly
relative to each other, 
but, on the other hand, short
distance effects also remain relevant because of the normal
divergences of the relativistic field theory.  In other words,
particle scattering near threshold is one situation where it is
important that the trial action contains information about both the
infrared and the ultraviolet behaviour of the theory (another
application where this would be important is in a confining theory
like Quantum Chromodynamics; see~\cite{Fey2}).  In the variational approach 
under discussion it is once again the non-locality in the proper time which
allows one to incorporate the physics of both these regimes into the
trial action simultaneously through suitable behaviour of the
retardation function for small and large values of $t_1-t_2$.

It is the purpose of the present paper to examine to what extent this
can be done for scattering, both analytically and numerically.  
A precursor to this
work may be found in Ref. (III).  There scattering was discussed at
{\it zeroth} order in the variational approximation (i.e. $S_{\rm eff}$ was
just replaced by $S_t$).  Although a drastic approximation, resulting
in a Gaussian dependence on all external momenta, the information
contained in the zeroth order variational scattering amplitude was
already considerable: for example, the forward scattering amplitude
had an analytic structure which contained the cuts associated with the
production of an arbitrary number of mesons in the intermediate state.
In fact, this is another general feature of this variational method:
the untruncated Green functions explicitly contain parts of {\it all}
Feynman diagrams and their associated analytic structure of any order
in the coupling constant (see (IV)).  Furthermore, it is possible to
perform the non-perturbative truncation of this Green function
analytically, yielding a rather concise formula for Green functions
involving an arbitrary number of external mesons.  In that paper, the
vertex function of the theory ( $n$ = 1) was examined in detail and it 
was found
that there are large non-perturbative effects, particularly near the
pair production threshold. In the present paper we generalize the variational
{\it ansatz} for the trial action further without modifying the variational
equations for the on-shell particles. Thus the vertex function remains 
unchanged but the variational scattering amplitude ( $n$ = 2 ) gets more
freedom to approximate the true off-shell propagation.

As in (III), we shall again pay particular attention to the question
of unitarity near the lowest threshold of the forward scattering
amplitude.  Because unitarity is not automatic in the formalism, it
serves as a useful guide to its success.  At zeroth order, unitarity
is already satisfied to within a factor of two.  We shall find that,
as expected, this is improved in the present calculation (i.e. at
`first order' in the variational calculation).  Guides such as this
are particularly important for the theory under consideration as the
Wick-Cutkosky model does not have a strong coupling
limit~\cite{Bay,paper1} and so it is difficult to assess the quality
of the variational calculation in the regime in which perturbation
theory fails \footnote{This should not be taken to mean that there are no
interesting non-perturbative effects even at modest couplings -- the
qualitative and quantitative differences between the perturbative and
variational form factors near threshold observed in (IV) are but one of
these.}.

This paper is organized in the following manner.  In the next Section
we describe the variational approach. In Section~3 we give the result
for the general ($2+n$)-point function.  We use this result in Section~4
to derive the scattering amplitude for two external mesons (i.e. $n=2$)
as a general function of the Mandelstam variables $s$, $t$ and $u$ and
examine its general structure as well as its region of convergence.
In Section~5 we specialize to the s-channel near the first threshold
and perform the necessary analytic continuation.  Furthermore, we
derive here important analytical results for the imaginary part of the
scattering amplitude near threshold and in Section~6 we numerically
investigate the question of unitarity.  Finally, in Section~7 we
conclude.

\section{ The Variational Approach}
\label{sec: Variational}
\setcounter{equation}{0}

The most general solvable trial action is of the form
\be
S_t[x] = \int_0^{\beta} \> dt \> \frac{1}{2}\dot{x}^2 + \int_0^{\beta}\>
dt_1 \int_0^{\beta} dt_2\> f(t_1-t_2)\> [x(t_1) -
x(t_2)]^2 \quad, 
\label{S_t}
\ee  
where $f(t_1-t_2)$ can be thought off as a `retardation function' 
depending on the proper time lapse $t_1-t_2$ between emission 
and re-absorption of mesons by the nucleon.
    
 In actual calculations it is preferable  to work in Fourier space
and write the trial action in the form
\be
S_t[b]= \sum _{k=0}^{\infty} A_k b_k^2 
\label{S_t[b]}
\ee
where the $b_k$ are the Fourier components of the nucleon path
\begin{equation}
x(t) = x \> \frac{t}{\beta} + \sum_{k=1}^{\infty} 
\frac{2 \sqrt{\beta}}
{k \pi} \> b_k \> \sin\left ( \frac{k \pi t}{\beta} \right ) \>
\quad.
\label{Fourier parameterization of paths}
\end{equation}
Defining $b_0=x/\sqrt{2\beta}$, the free action is simply
$S_0=\sum_{k=0}^{\infty} b_k^2$ and the nucleon propagator is given by
\bea
G_2(x) & = &\frac{1}{8\pi^2} \int_0^{\infty} \> d\beta \frac{1}{\beta ^2} 
\exp \left(- \frac{\beta}{2} M_0^2 -\frac{x^2}{2\beta} \right ) 
\int {\cal D} b \> e^{-S_{\rm eff}}       \\ \nonumber
& \ge&  \frac{1}{8\pi^2} \int_0^{\infty} \> d\beta \frac{1}{\beta ^2} 
\exp\left(- \frac{\beta}{2} M_0^2 -\frac{x^2}{2\beta}\right )
\exp(-<\Delta S>_{S_t}) \frac{\int {\cal D} b \> e^{-S_t}}{\int {\cal D}b \>
e^{-S_0}}
\label{ G_2(x)}
\eea
It is well known that near the pole it is the Fourier transform of
$G_2(x)$ which has a simple form.  For this reason it is advantageous to
apply the variational principle directly to 
$G_2(p)=\int d^4 x \>\exp(ip \cdot x)\> G_2(x)$ rather than
to $G_2(x)$ using a trial action which is also a function of $p \cdot x$. 
In (I)  a simple trial action  was constructed for the propagator
by  allowing an additional 
variational parameter $\lambda$ which can
be thought off as a rescaling of the momentum $p$ or a `velocity' parameter.
The form used was
\be
\tilde{S}_t= S_t \> -\> i \lambda p \cdot x
\label{S_t momentum} 
\ee
It was shown in (I) that in order to recover the correct pole behaviour
of the propagator the proper time $\beta$ must tend to infinity. In this
limit all discrete sums over the Fourier modes $A_k$ turn into integrals
over the profile function $A(E=k\pi/\beta)$. The `profile function' 
$ A(E) $ is linked to the retardation function $f(\sigma)$ through
\be
A(E)= 1 + \frac{8}{E^2} \int_0^{\infty} \> d\sigma \> f(\sigma)\> {\rm
sin}^2 \frac{E\sigma}{2} \quad.
\ee

The functional form of the retardation function, $f(t_1-t_2)$,
may be left free and is best determined using the variational principle.
In practice, however,
it is more convenient to use a particular parameterization containing a few 
variational parameters rather than a variational functional. This allows 
one to obtain analytic rather than numerical results. In our present
numerical scheme it is also necessary for performing the analytic continuation
from Euclidean time (where the variational calculation is done) to
Minkowski time (where the scattering amplitude is evaluated).
For the nucleon propagator it was found in  (II) that it was quite sufficient 
to use trial actions which have a similar behaviour
to the true action for small time intervals.
For scattering, especially near particle production thresholds, we also 
need the correct infrared behaviour of the theory. 
A particular example of a quantity which is sensitive to large distance 
scales is the imaginary part of a forward scattering amplitude near a 
threshold.  It must rise linearly with momentum 
because of the limited phase space available to the total
cross section which corresponds to it via the optical theorem. In Ref. (III)  
we used progressively more refined 
profile functions in order to investigate this. It was indeed found that
whether or not the variational forward amplitude  
exhibited this linear rise (at zeroth order) was critically dependent on 
the infrared behaviour of the associated trial action.
 The variational principle came to our aid by providing the best form for the 
retardation function for all values of $\sigma$ and therefore we could 
determine what the small as well as the large $\sigma$ behaviour of 
$f(\sigma)$ should be.
In (III) it was shown that one can incorporate the correct large- and 
small-$\sigma$ behaviour by using the following functional form for the 
retardation function (`extended' parameterization):
\be
f(\sigma) = \frac{C_1}{\sigma^2} \left [ e^{-w_1\sigma} - C_2
\sqrt{\sigma} e^{-w_2\sigma} \right] \quad,
\label{f extended}
\ee
with the requirement that $w_1>w_2$ and $C_2>0$. The associated profile
function is 
\bea
A(E) &=& 1 + \frac{2C_1}{E^2}\biggl [  E\>{\rm arctan}\frac{E}{w_1} -
w_1\>{\rm ln}\left(1+\frac{E^2}{w_1^2}\right)  \\ \nonumber
 & &\hspace{2cm} -4C_2\sqrt{\pi}  \left (
\sqrt{\frac{1}{2}\left(\sqrt{w_2^2+E^2}+w_2 \right)} - \sqrt{w_2}\right)
\biggr ] 
\label{ A extended}
\eea
having branch points at $E = i w_2$ and $E = i w_1$.
The variational parameters $C_1, C_2, w_1$ and $w_2$ were determined
from the variational equations for the 2-point function $G_2(p)$ with
the trial action of Eq.~(\ref{S_t momentum}) following the procedure
described in (I) and (II).

\section{ The ($2+n$) point function }
\label{sec:(2+n)}
\setcounter{equation}{0}

\begin{figure}[hbtp]
\begin{center}
\mbox{\epsfxsize=12cm\epsfysize=8cm\epsffile{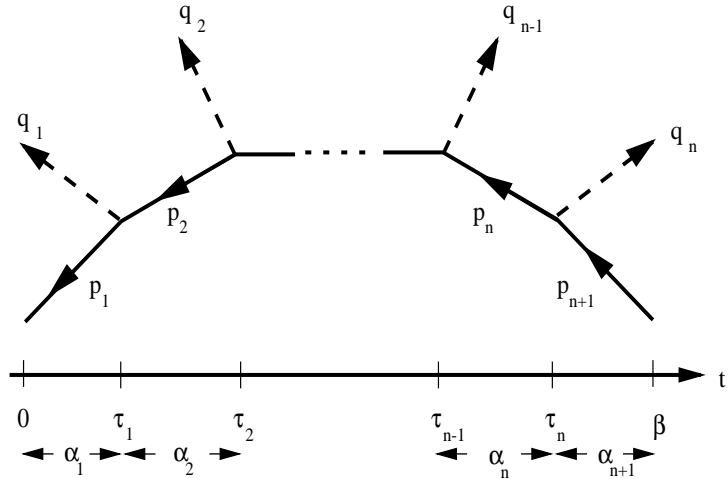}}
\end{center}
\vspace*{-1.5cm}
\caption{The definition of the relative times and momenta. Note 
that $\tau_0 = 0$, $\tau_{n+1} = \beta$, $p_1 = - p$ and 
$p_{n+1}=p'$.}
\label{fig:relative}
\end{figure}

The $(2+n)$-point function can be derived in a similar way to the
propagator $G_2(p)$. In (IV) it was shown that
the variational expression  for the $(2+n)$ function is given by
\be
G_{2,n}(p,p';\{q\}) \propto \int_0^\infty d\beta
\left [\prod_{i=1}^{n}  g \int_{0}^{\beta} d\tau_i\right ]  
\exp \left [ - {\beta \over 2} M_0^2 \right ] 
e^{-< \Delta \tilde S >_{\tilde S_t} } 
\int d^4x \int {\cal D} x e^{-{\tilde S_t}[x(t)]} \> .
\label{greenvar}
\ee  
In terms of the Fourier components $b_k$ the effective action in
momentum space 
is
\begin{equation}
\tilde S = S[x(t)] + i\left [ \> \sqrt {{2\over \beta}} b_0
\cdot \sum_{i=1}^{n+1} \alpha_i p_i - \sum_{k=1}^{\infty} 
Q_k^{(n)}\cdot b_k \> \right ]
\label{tilde s fourier}
\end{equation}
where
\begin{equation}
Q_k^{(n)} = {2 \sqrt {\beta} \over \pi k} \sum_{i=1}^{n}\sin 
\left ( {k \pi \tau_i \over \beta} \right ) q_i\;\;\;,
\end{equation}
These additional terms on the right hand side (or, rather, their
exponential) are just the
plane waves appropriate for the $n$ external mesons.  The Euclidean momenta 
and the integration variables $\alpha_i$ are defined
in Fig.~\ref{fig:relative}. \footnote{
The unrestricted integration over all $\tau_i$ in 
Eq.~(\ref{greenvar}) may be replaced by an integration over $\tau_i$ as 
indicated in Fig.~\ref{fig:relative}, provided that all permutations of
the meson momenta are summed over. In this way {\it one} wordline expression
contains {\it many} different Feynman diagrams }

\unitlength1mm

Given a true action of the above form, it was a natural choice in (IV)
to take the trial action for the n--point function to be of the form
\begin{equation}
\tilde S_t = S_t + i \left [ \> \sqrt {{2\over \beta}}\lambda  A_0 \, b_0
\cdot \sum_{i=1}^{n+1} \alpha_i p_i  - \sum_{k=1}^{\infty} \lambda A_k \,
\> Q_k^{(n)}\cdot b_k \right ]\;\;\;,
\label{general trial action1}
\end{equation}
where $\lambda$ and $A_k$ are variational parameters which
turned out to  have the same variational equations independently of
the number of external mesons $n$. \footnote{The reason why the combination
$\lambda A_k$ appears in the last term of Eq.~(\ref{general trial action1})
is connected to the truncation of the Green function -- see Ref. (IV).}

Actually, one can do somewhat better than this without much effort and,
indeed, it turns out to be physically sensible to choose a somewhat more
general trial action than Eq.~(\ref{general trial action1}).  The reason
is the following: 
although we only consider on-shell Green functions, 
the intermediate nucleon lines in Fig.~\ref{fig:relative} are in 
general still off-shell. Therefore
it might be useful to associate a different $\lambda$ with
each of the $(n+1)$ nucleon momenta in this action, thus allowing the
trial action additional freedom to adjust itself to the fact that these 
internal
lines have different virtualities.  As we shall see, for the correct
position of particle production thresholds in the intermediate state
this choice of trial action will prove to be useful. 

In the following, we briefly summarize the relevant results for a trial
action of this form.  The discussion follows (IV) closely, where the 
interested reader 
may find further details.  Crucially, as we shall see shortly, the
variational equation for the $\lambda$'s associated with the external legs
reduce to the same equation as one found for the nucleon propagator,
resulting in the same mass renormalization as before.
Therefore, a consistent truncation of the Green function remains possible.

\noindent
In short, the trial action which we shall use is given by
\begin{equation}
\tilde S_t = S_t + i \left [ \> \sqrt {{2\over \beta}} A_0 \, b_0
\cdot \sum_{i=1}^{n+1} \alpha_i p_i \lambda_i - \sum_{k=1}^{\infty} A_k \,
\tilde Q_k^{(n)}\cdot b_k \right ]\;\;\;,
\label{general trial action}
\end{equation}
where
\begin{equation}
\tilde Q_k^{(n)} = {2 \sqrt {\beta} \over \pi k} \sum_{i=1}^{n}\sin 
\left ( {k \pi \tau_i \over \beta} \right ) (\lambda_{i+1} p_{i+1} -
\lambda_{i} p_{i})\;\;\;,
\end{equation}
and $S_t$ is given by Eq.~(\ref{S_t[b]}).
Note that in the limit in which the coupling goes to zero the true 
action $\tilde S$ becomes
\begin{equation}
\tilde S_0 = S_0 + i \left [ \>
\sqrt {{2\over \beta}} b_0
\cdot \sum_{i=1}^{n+1} \alpha_i p_i - \sum_{k=1}^{\infty}  
Q_k^{(n)}\cdot b_k \> \right ]\;\;\;,
\end{equation}
so that $A_k$ and $\lambda_i$ must go to 1 in this limit.
Writing the true action as
\begin{equation}
\tilde S = \tilde S_0 + S_1
\end{equation}
one obtains for the path integral of the exponential of the trial
action, up to irrelevant overall constants, 
\begin{equation}
\int {\cal D}{\tilde x} e^{-\tilde S_t} = \left [
\prod_{k=0}^{\infty} {1 \over
A_k^2}\right ] \> \exp \left \{ -{1 \over 2 \beta} \left [
A_0 \tilde P^2 + {\beta \over 2} \sum_{k=1}^{\infty} A_k 
{\tilde Q_k^{(n)^2}} \> \right ] \> \right \}\;\;\;.
\end{equation}                     
where 
\be
\int{\cal D}\tilde{x} = \int d^4x \int_{x(0)=0}^{x(\beta)=x} {\cal D}x \> .
\ee
Further, as part of $<\Delta S>$ we need
\begin{equation}
<\tilde S_0 - \tilde S_t >_{\tilde S_t} =
{P \cdot \tilde P \over \beta} - {\tilde P^2 \over  2\beta} (1+A_0)
 + 2 \sum_{k=0}^{\infty} \left ( {1 \over A_k} - 1 \right )
- \sum_{k=1}^{\infty} { {\tilde Q_k^{(n)^2}} \over 4}
 (1+A_k)\>  + \> \sum_{k=1}^{\infty} {\tilde Q_k^{(n)} \cdot Q_k^{(n)}  
\over 2},  
\end{equation}
For brevity we have defined the quantity $P$ to be sum over all
the nucleon lines' momenta weighted by the respective proper time
intervals, i.e.
\begin{equation}
P = \sum_{i=1}^{n+1} \alpha_i p_i \;\;\;.
\end{equation} 
$\tilde P$ is the equivalent quantity containing the relevant factors of
$\lambda_i$, namely
\begin{equation}
\tilde P = \sum_{i=1}^{n+1} \alpha_i p_i \lambda_i \;\;\;.
\end{equation}  
Combining these terms we obtain
\begin{equation}
\exp \left [ - <\tilde S_0 - \tilde S_t>_{\tilde S_t} \right ] \> 
\int  {\cal D} \tilde{x} \> e^{- \tilde S_t}  =
\exp \left [ \> 2 \sum_{k=0}^{\infty} \left (1-{1 \over A_k} - 
\log A_k \right ) + \sum_{i=1}^{n+1} 
\alpha_i p_i^2 {\lambda_i (\lambda_i -2 ) \over 2} \> \right ] \quad.
\end{equation}
Finally, the weighted average of the interacting part of the
action is given by
\begin{equation}
< \> S_1 \> >_{\tilde S_t} = \> - \frac{g^2}{16 \pi^2} \> 
\int_0^{\beta} dt_1 dt_2 \> \frac{1}{\tilde \mu^2(\sigma,T)}
\int_0^1 dx_1 \> \> e\left ( m \tilde \mu(\sigma,T),\>
\frac{- i   \tilde W^{(n)}}{ \tilde \mu(\sigma,T)}\> , \> x_1
\right ) 
\label{tilde Fourier average S1 explicit'}
\end{equation}
where $e(x,y,z)$ is defined to be the exponential
\be
{\rm e}(x,y,z) = \exp \left ( \> -\frac{x^2}{2} \> \frac{1-z}{z} - 
\frac{y^2}{2} z \> \right ) \quad.
\label{e(x,y,z)}
\ee
The quantity $\tilde \mu^2(\sigma, T)$ appearing in the
above equations is defined in terms of the 
`pseudotime' $\mu^2(\sigma,T)$, i.e. 
\begin{eqnarray}
\tilde \mu^2(\sigma,T) \> &=&\>  \frac{\sigma^2}{A_0 \beta} \>
+ \> \mu^2(\sigma,T) \nonumber \\
 &=&\>  \frac{\sigma^2}{A_0 \beta} \> +{ 8 \beta \over \pi^2}
 \> \sum_{k=1}^{\infty} \frac{1}{A_k \> k^2} \> 
\cos^2 \left(\frac{k \pi T}{\beta} \right )  \>
\sin^2 \left(\frac{k \pi \sigma}{2 \beta} \right ) \> .
\label{tilde amu2}
\end{eqnarray}
Here  we have introduced the relative and total times 
$\sigma = t_1 - t_2$ and $T = (t_1 + t_2)/2$
and the pseudotime $\mu^2$, after taking the $\beta \rightarrow \infty$
limit, becomes 
\begin{equation}
\mu^2(\sigma) \>  =
\>  \frac{4}{\pi} \int_0^{\infty} dE \> \frac{1}{A(E)} \>
\frac{\sin^2 (E \sigma/ 2)}{E^2} \>.
\label{amu2(sigma)}
\end{equation}
The analytic structure as well as 
the asymptotic behaviour of the pseudotime in Eq.~(\ref{amu2(sigma)})
is of great importance.  This analytic structure is
closely linked to the analytic structure of the profile function.
In particular, parameterizations of the profile function
which exhibit the same ultraviolet behaviour as the true action
lead to a pseudotime $\mu^2 (\sigma)$ which grows linearly with
$\sigma$ for both $ \> \sigma \to  0 \> $ and $ \> \sigma \to  \infty \> $ 
and does not have poles or cuts in the half-plane where 
${\cal R}e \> (\sigma)$ is positive (see III).

\noindent
The quantity $\tilde W^{(n)}$ is defined by 
\begin{equation}
\tilde W^{(n)}={\sigma \over \beta} \tilde P - {2 \sqrt{\beta} \over \pi}
\sum_{k=1}^{\infty} \tilde Q_k^{(n)} \frac{1}{k} 
\cos \left(\frac{k \pi T}{\beta} 
\right ) \> \sin \left(\frac{k \pi \sigma }{2 \beta} \right ) \> .
\label{eq:Wn Fourier}
\end{equation}
It is equal to
\begin{equation}
\tilde W^{(n)}=  {\sigma \over 2 } (\lambda_1\> p_1 + \lambda_{n+1}\> 
p_{n+1}) + {1 \over 2} 
\sum_{i=1}^{n} \> \bigl ( \> \left | \tau_i - t_1 \right | -  
\left | \tau_i - t_2 \right | \> \bigr ) \> (\lambda_{i+1}\> p_{i+1} - 
\lambda_{i}\> p_{i})  
\label{eq:Wn abs}
\end{equation}
which simplifies to
\begin{eqnarray}
\tilde W^{(n)} &=& {\rm sign} (\sigma) \Bigl \{\> \biggl [ \> \tau_{b+1} - 
\min(t_1,t_2) \> \biggr ] \> \lambda_{b+1}\> p_{b+1} 
\label{eq:wn} \\
&&\hspace{1cm} + \sum_{i=b+2}^{a-1} \alpha_i \lambda_i p_i
+ \> \biggl [ \> \max(t_1,t_2) - \tau_{a-1} \> \biggr ] 
\> \lambda_{a}\> p_{a} \> \Bigr \} \hspace{1cm} a > b + 1 \nonumber \\
&=& \sigma \lambda_{a} p_{a} \hspace{8.7cm} a = b + 1 \> .
\nonumber
\end{eqnarray}
For the case $a = b + 2 $, the sum appearing in Eq.~(\ref{eq:wn}) is 
defined to be empty. In other words, $\tilde{W}^{(n)}$ is equal  
(up to a sign, which is not relevant as only the
square of $\tilde{W}^{(n)}$ enters) to the integral of the proper time 
multiplied by the modified nucleon's four-momentum $\lambda p$ for the 
duration of the exchange of the internal meson. 
Particular cases of interest, for the purpose of the eventual 
truncation of the Green function, are those where $t_{1,2}$
are both smaller or both larger than all $\tau_i$,  in which case
$\tilde{W}^{(n)}$ $ = $ $\sigma \lambda_1 p_1$ and 
$\sigma \lambda_{n+1} p_{n+1}$, respectively.

Having calculated the above averages, one can now write down the expression 
for the (untruncated) $(2 + n)$-point function to first order in the 
variational framework
\begin{eqnarray}
G_{2,n}(p,p';\{q\})&=& {N_0 \over 2 g}  \sum_{{\cal P}\{q_i\}} 
\left \{ \prod_{i=1}^{n+1} g \> \int_0^\infty d\alpha_i \>   
\exp \left [ - {\alpha_i \over 2} 
\left ( M_0^2 + 2 \Omega + p_i^2 [ 1 - (1 - \lambda_i)^2 ] \> \right )
\right ] \>  \right \}
\label{eq:unrenormalized}\\
& &\hspace{0.3cm} \cdot \>  
\exp \left [ \frac{g^2}{8 \pi^2} \> \int_0^{\beta} dt_1
\int_{0}^{t_1} dt_2 \> {1 \over \mu^2(\sigma)}
\int_0^1 dx_1 \> \> e\left ( m  \mu(\sigma),\>
\frac{- i \tilde  W^{(n)}}{\mu(\sigma)}\> , \> x_1
\right )\right ] \>.\nonumber
\end{eqnarray}
Here $N_0$ is defined to be
\begin{equation}
N_0 = {1 \over A(0)}\>\exp \left ( 1 - \frac{1}{A(0)} \right ) \> .
\end{equation}
For $i=1$ and $i=n+1$ we get the same on-shell relation between the bare 
mass and the renormalized mass as before, e.g. for $i=1$
\begin{equation}
M_0^2 + 2 \> \Omega = M^2 [ 1 - (1 - \lambda_1)^2 ] + 
\frac{g^2}{4 \pi^2} \> \int_0^{\infty} { d\sigma \over \mu^2(\sigma)}
\int_0^1 dx_1 \> \> e\left ( m  \mu(\sigma),\>
\frac{\lambda_1  \sigma M}{\mu(\sigma)}\> , \> x_1 \right )
\label{eq:MM0}
\end{equation}
where the quantity $\Omega$ has been defined in Ref. (I).
Further, we get the same variational equations for these two lambda's as 
before and so we write
\begin{equation}
\lambda_1 \> = \> \lambda_{n+1} \> = \>\lambda
\end{equation}
The truncated $(2 + n)$-point function 
$G_{2,n}^{tr}(p,p';\{q\})$, valid for on-shell external nucleon 
momenta, now reads
\begin{eqnarray}
G_{2,n}^{\rm \> tr}(p,p';\{q\})\>&=&\> {2\> g \over N_0} \>  
\sum_{{\cal P}\{q_i\}} 
\left \{ \> \prod_{i=2}^{n} g \> \int_0^\infty d\alpha_i  \> 
\exp \Bigl [ \> - {\alpha_i \over 2} \left (
  M^2 ( 1 - (1 - \lambda)^2 ) + p_i^2 ( 1 - (1 - \lambda_i)^2 )  \> 
\right ) \> \Bigr ]\> \right \} \nonumber \\
& &\hspace{1.5cm}  \cdot \>
\exp \Biggl \{ - \frac{g^2}{8 \pi^2} \>  \int_0^1 du \>\biggl [
\int_0^{\infty} {d\sigma  \over \mu^2(\sigma)} 
\sum_{i=2}^{n}\alpha_i \> e\left ( m  \mu(\sigma),\> 
\frac{\lambda \sigma M}{\mu(\sigma)}\> , \> u \right ) \nonumber \\
& &\hspace{2.5cm}
- \> \int   {dt_1 dt_2 \over \mu^2(\sigma)} \>
 e \left ( m  \mu(\sigma),\> \frac{- i  
\tilde W^{(n)}}{\mu(\sigma)}\> , \> u \right ) \biggr ] \> \> \>\Biggr \} \>.
\label{eq:truncated}
\end{eqnarray}
where ${\cal P}\{q_i\}$ denotes all possible permutations of $q_i$.
Eq.~(\ref{eq:truncated}) is the main result of this section.

\section{ The (2+2) scattering amplitude}
\label{sec: scatt. ampl.}
\setcounter{equation}{0}

\begin{figure}[hbtp]
\begin{center}
\mbox{\epsfxsize=8cm\epsfysize=8cm\epsffile{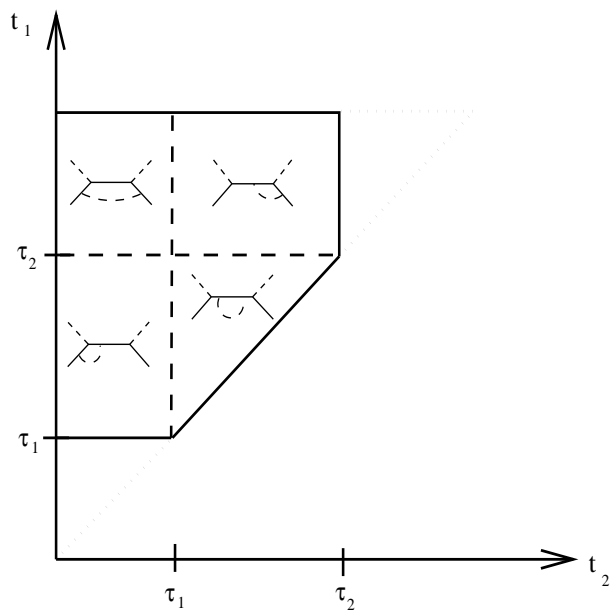}}
\end{center}
\caption{The $(t_1$ -- $t_2)$ - integration region is shown enclosed
by the solid 
line. The (dotted) triangular regions are those relevant
 to the dressing of the external legs and have been removed
in the truncated amplitude.  Also shown are the relevant diagrammatic
representations (for the direct diagram) of the {\it action} in the 
particle representation}
\label{fig:region}
\end{figure}

In this section we focus our attention to the description of the
scattering amplitude for meson-nucleon scattering. From the last
equation of the previous section, substituting $n = 2$, we immediately
obtain
\begin{eqnarray}
G_{2,2}(p,q_1,q_2)&=& {2\> g^2 \over N_0} \>  \sum_{{\cal P}\{q_1,q_2\}} 
 \> \int_0^\infty d\alpha_2 \> \exp \left \{ 
- {\alpha_2 \over 2} \left [
  M^2 ( 1 - (1 - \lambda)^2 ) + p_2^2 ( 1 - (1 - \lambda_2)^2 )  \> 
\right ] \right. \nonumber \\
&& \hspace{2cm}\left.
+ \> {g^2 \over 8 \pi^2} \> \tilde \Xi (p,q_1,q_2;\alpha_2) \right \} \quad.
\label{eq:truncated22}
\end{eqnarray}
The function $\tilde \Xi$ is defined as
\begin{eqnarray}
 \tilde \Xi (p,q_1,q_2;\alpha_2) &=&\>  \int_0^1 dx_1 \>\left \{
 \int   {dt_1 dt_2 \over \mu^2(\sigma)} \>
 e \left ( m  \mu(\sigma),\> \frac{- i \tilde W^{(2)}}{\mu(\sigma)}\> , 
\>x_1 \right ) 
\right.  \nonumber \\
& &\hspace{2.5cm} - 
\left. \int_0^{\infty} {d\sigma  \over \mu^2(\sigma)} \> 
\alpha_2 \> e\left ( m  \mu(\sigma),\> 
\frac{\lambda \sigma M}{\mu(\sigma)}\> , \> x_1 \right )\right \} \> 
\> \> \>.
\label{eq:Xidef}
\end{eqnarray}
The $t_1 - t_2$ integration region  in Eq.~(\ref{eq:Xidef}) is shown in 
Fig.~\ref{fig:region}.  Depending on the ordering of the 
internal meson's emission and absorption  times $t_2$ and $t_1$ with
respect to the proper times at which the external mesons interact with
the nucleon, i.e. $\tau_1$ and $\tau_2$, the function
$\tilde{W}^{(2)}$ assumes a different value.  Explicitly,
\begin{eqnarray}
\tilde W^{(2)}=& (\tau_1 - t_2) \lambda \> p_1 + (t_1 - \tau_1)\lambda_2 \>  
p_2\hspace{2cm} & 
t_2 < \tau_1\;\;, \tau_1 < t_1 < 
\tau_2 \nonumber \\
=& (t_1 - t_2) \lambda_2 \> p_2& \tau_1 < t_1,t_2 < \tau_2 \nonumber \\
=& (\tau_2 - t_2) \lambda_2 \> p_2 + (t_1 - \tau_2)\lambda \>  p_3 & 
\tau_1 < t_2 < \tau_2\;\;, \tau_2 < t_1 \nonumber \\
=& (\tau_1 - t_2) \lambda \> p_1 + \alpha_2 \lambda_2 \> p_2 + (t_1 - \tau_2) 
\lambda \> p_3 & 
t_2 < \tau_1 \;\;,\tau_2 < t_1 \> . 
\label{eq:Wn abs2}
\end{eqnarray}

\unitlength1mm

Although the integration variable $\alpha_2$  in the expression for the Green 
function in Eq.~(\ref{eq:truncated}) is a time difference, the integration
variables $t_1$ and $t_2$ in Eq.~(\ref{eq:Xidef}) are still absolute times.  
This leads
to a spurious dependence of $\tilde{W}^{(2)}$ and the integration limits of 
the $ (t_1$ -- $t_2) $ - integration on the absolute times $\tau_1$ and 
$\tau_2$ \footnote{It should be remembered that $\tau_1$ and $\alpha_3$ 
have already been integrated over in Eq.~(\ref{eq:truncated}) in order to 
truncate the Green function, so a dependence
on these variables in Eq.~(\ref{eq:truncated}) would be non-sensical.}. In
order to eliminate this spurious dependence it is convenient to make a
transformation of the integration variables $t_1$ and $t_2$ to two variables
which {\it are} expressed as time differences.
Clearly $\sigma = t_1 - t_2$ is a suitable choice for one of these.
It is convenient to define the second variable $x_2$ in a different
way for each of the integration regions depicted in Fig.~\ref{fig:region}.
We shall discuss each of these regions below.

Before proceeding, let us make the change
to the Minkowski space Mandelstam variables $s$, $t$ and $u$ (as well as
Minkowski $q^2$ and $q'^2$ for the external mesons), which are
defined in terms of the Euclidean momenta $p_1$, $p_3$, $q_1$ and $q_2$
through
\begin{eqnarray}
s &=& - (p_1 + q_1)^2 \nonumber \\
t &=& - (p_1 - p_3)^2 \\
u &=& - (p_1 + q_2)^2\;\;\;. \nonumber 
\end{eqnarray}
In contrast to the usual formalism
this change from Euclidean to Minkowski four-momenta is a trivial
one in the wordline approach, as there are no four-momentum integrals left 
to be done in the expression for the Green function.

It is rather useful to break up the function
$\tilde \Xi$ not only into the components corresponding to the two
vertex corrections, the propagator correction and the box diagram
depicted in Fig.~\ref{fig:region}, i.e.
\begin{equation}
\tilde \Xi(s,t,q^2,q'^2;\alpha_2)=  \tilde \Xi_{V}(s,q^2;\alpha_2) +  
\tilde \Xi_{V}(s,q'^2;\alpha_2) + \tilde \Xi_{P}(s;\alpha_2) 
+ \tilde \Xi_{B}(s,t,q^2,q'^2;\alpha_2)\;\;\;,
\end{equation}
but also to immediately extract their leading $\alpha_2$ -  behaviour.
The reason for this is that $\alpha_2$ is the proper time
conjugate to the intermediate nucleon's momentum and so one
might intuitively expect that the large $\alpha_2$ - behaviour is particularly
important close to the threshold, i.e. close to where real particles could
be produced in the intermediate state.  This is analogous to 
what happens for the nucleon propagator itself -- as indicated in Section
2, the on-shell propagator
corresponds to letting $\beta \rightarrow \infty$.  This physical
interpretation associated with the use of the Fock--Schwinger proper
time formalism makes it advantageous over the use of the usual Feynman 
parameters.

Explicitly, the component of $\tilde \Xi$ corresponding
to the bottom left hand integration region in Fig.~\ref{fig:region}
may be written as
\begin{equation}
\tilde \Xi_{V}(s,q^2;\alpha_2) = \Xi_{V}^0(s,q^2) + 
\Xi_{V}(s,q^2;\alpha_2)\;\;\;,
\label{eq:V1}
\end{equation}
where $\Xi_{V}^0(s,q^2)$ does not depend on $\alpha_2$ and 
$\Xi_{V}(s,q^2;\alpha_2)$ goes to
zero as $\> \alpha_2 \rightarrow \infty \> $.    
In these variables, the constant piece $\Xi_{V}^0(s,q^2)$ for the
direct diagram (we shall discuss the crossed diagram later)
becomes
\begin{equation}
\Xi_{V}^0(s,q^2) =  \int_0^1 dx_1 dx_2 \int_0^{\infty} d\sigma \> 
{\sigma \over \mu^2 (\sigma)} \> e\left ( m \mu (\sigma), 
{\sigma \over \mu (\sigma)}
  \> W_{V}(x_2),x_1 \right )\;\;\;.
\label{eq:v0}
\end{equation}
Here the variable $x_2$ is defined as the time interval $\tau_1 - t_2$ 
between the external and internal meson's interaction time with the nucleon,
scaled by $\sigma$, and $W_{V}^2(x_2)$ is given by
\begin{equation}
W_{V}^2(x_2) = \left [ \> x_2 \lambda_2 + {x_2}^2 ( \lambda - \lambda_2 )
\> \right ] \lambda M^2  \> + \> \lambda_2 ( 1 - x_2) \left [ \> 
s \left ( \lambda_2 + x_2 
(\lambda - \lambda_2) \right ) -  \lambda x_2 q^2 \> \right ] \> .
\end{equation}
On the other hand, the piece of $\tilde \Xi_{V}$ which depends on $\alpha_2$ 
is
\begin{equation}
\Xi_{V} (s,q^2;\alpha_2) = - \int_0^1 dx_1 dx_2 \int_0^{\infty} d\sigma \> 
{\sigma \over \mu^2 (\sigma + \alpha_2)} \> 
e\left ( m \mu (\sigma + \alpha_2), 
{(\sigma + \alpha_2)  \over \mu (\sigma + \alpha_2)}
\> W_{V}({\sigma \over \sigma + \alpha_2} x_2),x_1 \right ),
\label{eq:va}
\end{equation}
where $\sigma$ has been shifted by $\alpha_2$ and $x_2$ has been rescaled
accordingly.  Also, it is easily seen that the expression for the second vertex
(top right hand region in Fig.~\ref{fig:region}) is the same, except that
$q^2$ needs to be replaced by $q'^2$.

In a similar way to the vertex function discussed in (IV)
the representations Eqs.~(\ref{eq:v0}) and~(\ref{eq:va}) for  
 $ \> \tilde \Xi_{V}(s,q^2;\alpha_2) \> $ and 
$\> \tilde \Xi_{V}(s,q'^2;\alpha_2) \> $
do not necessarily converge for all $s$.  In fact, if we
restrict ourselves to the case where the external mesons are on-shell (i.e.
$q^2$ = $q'^2$ = $m^2$) then Eqs.~(\ref{eq:v0}) and~(\ref{eq:va}) are only
valid representations if $s$ $>$ $0$ .
\footnote{This requirement originates in the
fact that $\mu^2(\sigma)$ grows like $\sigma$ for large $\sigma$, so that
for the $\sigma$ - integral to converge we require $W_V^2$ $>$ $0$ for all
$0 \le x_2 \le 1$.} If one is interested in $s < 0$, then
Eqs.~(\ref{eq:v0}) and~(\ref{eq:va}) must be analytically continued through
the use of a contour integration in the complex $\sigma$ - plane.

\begin{figure}[hbtp]
\begin{center}
\mbox{\epsfxsize=12cm\epsfysize=8cm\epsffile{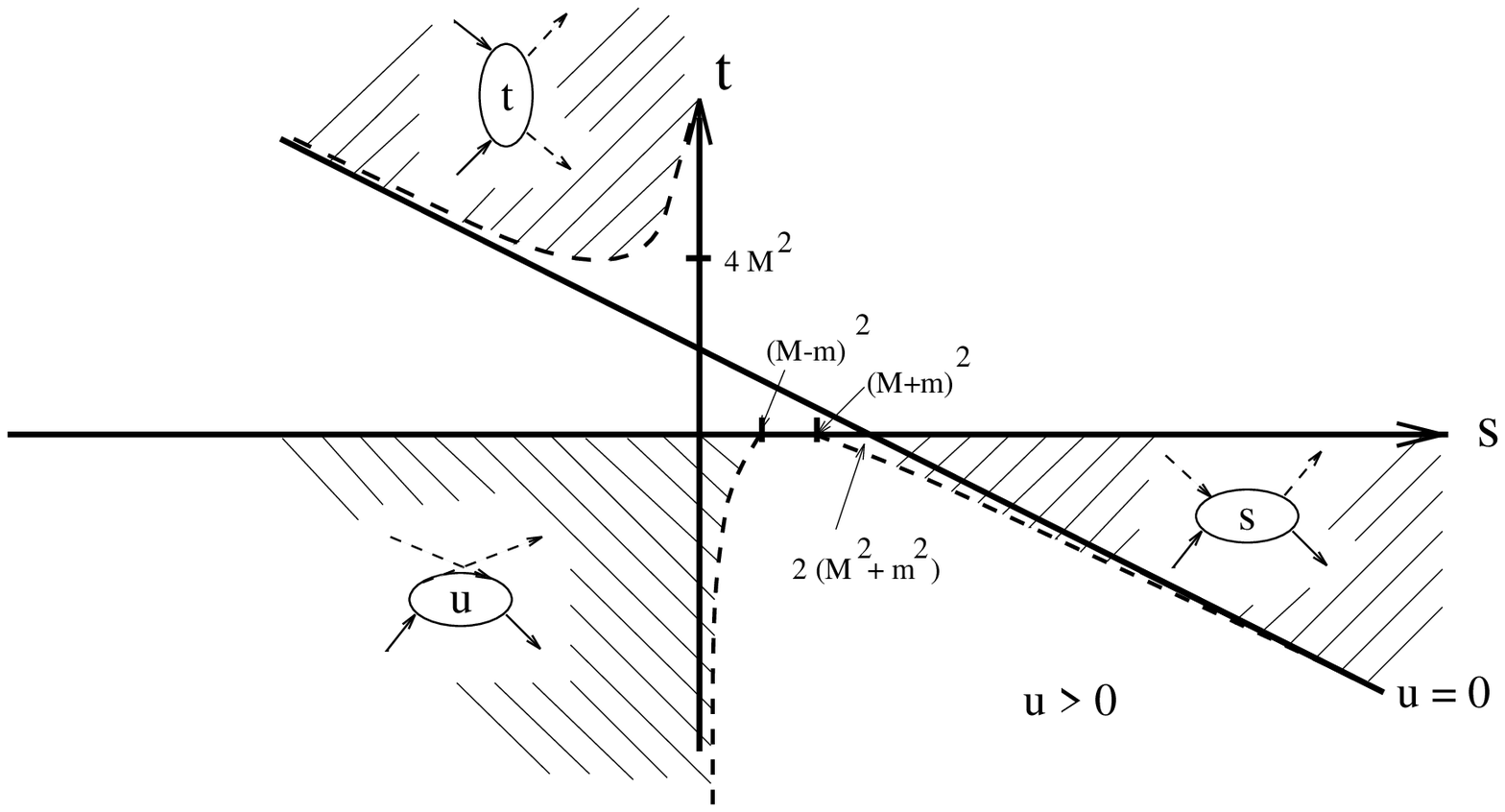}}
\end{center}
\caption{The Mandelstam plot for the scattering process}
\label{fig:mandelstam}
\end{figure}

It is worthwhile at this point to remind oneself of the corresponding 
situation for the vertex function (see Ref. (IV)). Here the original (real) 
representation
for $G_{2,1}(q^2)$  was valid for all values of $q^2$ below the physical
region for nucleon pair production (i.e. $q^2$ $<$ $4 M^2$).  For $q^2$
larger than this the same contour rotation as above had to be performed, 
resulting in a cut and hence an imaginary part for the
amplitude, starting at the nucleon pair production threshold. 
The situation for $\tilde \Xi_{V}$ is reversed, as
we have just indicated that this is real for $s$ $>$ $0$ (and thus
real in the $s$-channel -- see Fig.~\ref{fig:mandelstam}), while
the corresponding quantity seems to become complex in the $t$ and (most of
the) $u$ channel.
 At first sight this seems fatal -- after all, the contribution
from the direct diagram should have a cut beginning at $s$ = $(M+m)^2$
and should be purely real in the $u$ channel.  We shall address
this issue and its solution below, after having discussed the
contributions to $\tilde \Xi$ from the other two integration regions.

\unitlength1mm

The part of $\tilde \Xi$ originating from the propagator correction may be
decomposed in a similar manner to $\tilde \Xi_{V}$.  This time,
however, there is also a term which goes linearly with $\alpha_2$:
\begin{equation}
\tilde \Xi_{P}(s;\alpha_2) = 
- \alpha_2 \> \Xi_{P}^1(s) + \Xi_{P}^0(s) + \Xi_{P}(s;\alpha_2) \> .
\label{eq:P}
\end{equation}
There is a good reason for the existence of this linear term for this
particular integration region; if we would expand our expressions in
a perturbation series (so that the diagrams in Fig.~\ref{fig:region}
would actually correspond to Feynman diagrams), this term would be 
responsible for the self-energy correction $\Sigma (s)$ of the intermediate 
off-shell propagator.  Mathematically, the term linear in $\alpha_2$
results from the fact that the integration region in the complementary
variable to $\> \sigma$ (which may be performed analytically) grows like 
$\alpha_2$, as may be seen in 
Fig.~\ref{fig:region}.   Clearly this is the only region for which this 
happens.  It is also no accident that the expression
for $\Xi_{P}^1$ looks very similar to those encountered  in the discussion
of the renormalization of the nucleon propagator in (I) 
\begin{equation}
\Xi_{P}^1(s) =  \int_0^1 dx_1  \int_0^{\infty}
{d\sigma \over \mu^2 (\sigma)}
\left \{
e\left ( m \mu (\sigma), 
{\sigma \over \mu (\sigma)}
\lambda  M ,x_1 \right )
-e\left ( m \mu (\sigma), 
{\sigma \over \mu (\sigma)}
\lambda_2  {\sqrt s} ,x_1 \right )
\right \}\;\;\;,
\label{eq:P1}
\end{equation}
the only difference being that  there it only applied to the on-shell 
propagator, while  $ \> \Xi_{P}^1 \> $ also contains off-shell information.
Indeed, as was proven for the general $n$-point function in (IV), the
variational calculation -- if expanded in the coupling -- reproduces 
perturbation theory to
one loop order exactly, {\it including} the off-shell behaviour of the
internal propagator. 

\noindent
 The constant piece contributing to
$\tilde \Xi_P$ is given by
\begin{equation}
\Xi_{P}^0(s) =  - \int_0^1 dx_1  \int_0^{\infty} d\sigma \> 
{\sigma \over \mu^2 (\sigma)} \> 
e\left ( m \mu (\sigma), 
{\sigma \over \mu (\sigma)}
\lambda_2  {\sqrt s},x_1 \right ),
\label{eq:P0}
\end{equation}
while the piece that vanishes as $\alpha_2 \rightarrow \infty$ becomes,
after shifting the $\sigma$ integration by $\alpha_2$ as before,
\begin{equation}
\Xi_{P} (s;\alpha_2) =  \int_0^1 dx_1  \int_0^{\infty} d\sigma \> 
{\sigma \over \mu^2 (\sigma + \alpha_2)} \> 
e\left ( m \mu (\sigma + \alpha_2), 
{\sigma + \alpha_2 \over \mu (\sigma + \alpha_2)}
\lambda_2  {\sqrt s}
,x_1 \right )\;\;\;.
\label{eq:Pa}
\end{equation}
Clearly, the region of convergence of the representation for
$\tilde \Xi_{P}$ (Eqs.~(\ref{eq:P}--\ref{eq:Pa})) is the same as that
of $\tilde \Xi_{V}$, i.e. $\> s > 0 \> $.  Also, it is relevant to the
subsequent discussion of 
the region of convergence of the $\alpha_2$ - integration that in the
$s$ channel $ \> \Xi_{P}^1 \> $ is positive.

Finally, the integration over the region where the action is represented by
the box diagram is given by
\begin{equation}
\Xi_{B} (s,t,q^2,q'^2;\alpha_2) =  
\int_0^1 dx_1 dx_2 \int_0^{\infty} d\sigma \> 
{\sigma \over \mu^2 (\sigma + \alpha_2)}
e\left ( m \mu (\sigma + \alpha_2), 
{\sigma + \alpha_2 
\over \mu (\sigma + \alpha_2)}   W_B ,x_1 \right )\;\;\;,
\end{equation}
where
\begin{equation}
W_B^2 = {(\sigma \lambda + \alpha_2 \lambda_2)
\over (\sigma + \alpha_2)^2} (\sigma \lambda M^2 + \alpha_2 \lambda_2 s) -
\left ({\sigma \lambda \over \sigma + \alpha_2}\right )^2(1-x_2) x_2 t
- {\sigma \alpha_2 \lambda \lambda_2 \over (\sigma + \alpha_2)^2}
\left [ x_2 q^2 + (1-x_2) q'^2 \right ] \;\;.
\end{equation}
This time there is not even a constant term in $\alpha_2$, and the
representation converges (for on-shell external mesons) as long as 
$t$ $<$ $4 M^2$, i.e. in the $u$ and $s$ channel.

In short, the contribution to the scattering amplitude due to
the direct diagram is given by
\begin{equation}
A_{2,2}^{\rm direct}(s,t,u,q^2,q'^2) = {\cal F}(s,q^2,q'^2) \int_0^\infty 
 \! \! \! d\alpha_2  
\exp \left \{ {\alpha_2 \over 2}  [ 1 - (1 - \lambda_2)^2 ]
[ s - \tilde s (s) ]
 + {g^2 \over 8 \pi^2}  \Xi (s,t,q^2,q'^2;\alpha_2) \right \}\;,
\label{eq:diramp}
\end{equation}
where
\begin{equation}
\tilde s(s) = \frac{1}{1 - (1 - \lambda_2)^2} \left[ \> (1 - (1 - \lambda)^2) 
M^2 \> + \> {g^2 \over 4 \pi^2} \> \Xi_P^1(s) \> \right] 
\end{equation}
and
\begin{equation}
{\cal F}(s,q^2,q'^2) = {2\> g^2 N_1 \over \lambda} \> 
\exp \left [ \> {g^2 \over 8 \pi^2}
\left ( \> \Xi_{V}^0(s,q^2) + \Xi_{V}^0(s,q'^2) + \Xi_{P}^0(s) \> \right )
\> \right ] \>.
\end{equation}
Furthermore $ \> N_1=\lambda Z/N_0  \> $ where $Z$ is the residue at the 
pole of the propagator.
Note that for reasonably small couplings $ \tilde s $ is `close' to $M^2$ and
that for $ \> s > M^2 \> $ it is larger than $M^2$.

Let us now come back to the question of the analytic structure of this
amplitude.  It is clear from Eq.~(\ref{eq:diramp}) that not only does one need
to address the convergence criteria of the $\sigma$ - integration, but also
simultaneously those of the $\alpha_2$ integration.  It is for this reason
that the analytic structure of the scattering amplitude is much richer than
that of the vertex function in (IV).  It is clear from Eq.~(\ref{eq:diramp})
that for the $\alpha_2$ - integration to converge we require 
$\> s <\tilde s \> $ (note that $\lambda$ is always less than 1).  However,
$\tilde s$, using the representation for $\> \Xi_P^1 \> $ in Eq.~(\ref{eq:P1}),
is only defined for $ \> s > 0 \> $.  So at the moment our Green function
is only defined in a region $ \> 0 <   s < \tilde s \> $ and 
$\> t < 4 M^2 \> $, i.e. 
except for a limited region of the $u$ -- channel 
(Fig.~\ref{fig:mandelstam}) it is only defined outside the three
physical regions for the process. Importantly, however, for the
analytic continuation there {\it does} exist {\it a}
region where it is defined.
Because of the complicated interplay between the convergence of the 
$\sigma$ -  and $\alpha_2$ -
integrations one needs to analytically continue the above representation on
a case by case basis.  In the present paper we are interested in 
doing this for the $s$--channel. 

Before we proceed, let us write down
the amplitude for the crossed channel.  It is given by
\begin{equation}
A_{2,2}^{\rm crossed}(s,t,u,q^2,q'^2)\>=\> 
A_{2,2}^{\rm direct}(u,t,s,q'^2,q^2)\;.
\end{equation}
Hence, again for on-shell external mesons, the representations above are only 
appropriate for $A_{2,2}^{\rm crossed}$
 in the region
$\> 0 < u <\tilde s \> $ and $\> t < 4 M^2 \> $.  As may be
seen in Fig.~\ref{fig:mandelstam}, this includes some of the physical region 
of the $s$ channel.  Indeed, as we shall concentrate our attention
on the region near the threshold at $s$ = $(M+m)^2$, this representation 
of  $A_{2,2}^{\rm crossed}$ suffices for our purposes.

\section{Unitarity in the $s$ channel near threshold}
\label{sec:s channel}
\setcounter{equation}{0}

In order to be able to describe scattering of on-shell mesons off the nucleon 
in the $s$ channel, the convergence requirement of the
$ \alpha_2 $ - integration of the direct diagram requires the
amplitude in  Eq.~(\ref{eq:diramp}) 
to be analytically continued so that it converges for $s > \tilde s$.
This may be done in an analogous manner to a Wick rotation as in 
Refs. (III,IV):
for $s$ infinitesimally above the cut, i.e. $s \rightarrow s + i \epsilon$,
one may deform the $ \alpha_2 $ - integration in such a way that it runs along
the positive imaginary axis.  This is possible because the contribution
of the quarter circle at infinity vanishes and because the analytic behaviour
of the pseudotime $\mu^2(\sigma)$ is known.  Changing variables to 
$\alpha_2 \rightarrow \alpha_2/i$, and suppressing the dependence 
on $q^2 = q'^2 = m^2$
for notational simplicity, one obtains
\begin{eqnarray}
A_{2,2}^{\rm direct}(s,t)\>&=&\> 
i {\cal F}(s)\int_0^\infty d\alpha_2 
\exp \left \{ \> i {\alpha_2 \over 2} [ 1 - (1-\lambda_2)^2 ]
\> [ s - \tilde s(s) + i \epsilon ] \>
+ \> {g^2 \over 8 \pi^2}  \> \Xi (s,t; i\alpha_2) \> \right \} 
\label{eq:foramp}\\
&=&\> {-2 {\cal F}(s) \over  [ 1 - (1 - \lambda_2)^2 ]\> [ s - \tilde s(s) 
+ i \epsilon] }
\>+\> i {\cal F}(s)\int_0^\infty d\alpha_2 \> 
\exp \left \{ i {\alpha_2 \over 2} [ 1 - (1 - \lambda_2)^2 ] 
[ s - \tilde s(s) ]  \right \}\nonumber\\
&&\hspace{7cm}\cdot \left \{ \> \exp \left [ \> {g^2 \over 8 \pi^2} \>  
\Xi (s,t; i\alpha_2) \> \right ] \> - \> 1 \> \right \} \nonumber\;\;\;,
\end{eqnarray}
where the latter expression, obtained by adding and subtraction a `modified 
Born term', does not require an $i \epsilon$ prescription for convergence 
and is therefore more suitable for numerical evaluation.

This expression for the amplitude converges in the region of interest.  
Together with the
crossed diagram, which does not have to be analytically continued as long as
one does not move more than about one nucleon mass away from the lowest 
threshold
(and therefore remains real), it yields the complete scattering amplitude.  
Furthermore, it should be noted that for the special case of forward 
scattering 
the expressions for $\tilde \Xi$ naturally simplify somewhat.  This is 
particularly true for the contribution of the box diagram to the action, 
where the $x_2$ - integration may
be carried out trivially when $t = 0$ and $q^2 = q'^2$.  Also, the
two vertex contributions are, of course, identical whenever the latter of these 
conditions is fulfilled.   

It is relatively straightforward to evaluate the crossed diagram 
numerically.  Both
the $\alpha_2$ - as well as the $\sigma$ - integral in 
$ \> \Xi (u,t; \alpha_2) \> $ converge exponentially, so simple Gaussian  
quadrature is quite adequate (we typically used 64, 72 or 96 Gaussian 
points per integration).  The direct diagram, however is much more difficult 
to handle as in this case the integrands
are oscillatory functions. In particular, the $ \alpha_2 $ -  integration 
only converges relatively slowly as $\alpha_2 \rightarrow \infty$.  
Moreover, the separate contributions to
$ \> \Xi(s,t;\alpha_2) \> $ arising from the vertex, propagator and 
box diagrams oscillate
with different frequencies for asymptotic $\alpha_2$, so that the numerical
behaviour of the exponential of $\Xi(s,t;\alpha_2)$ is rather erratic.
This means that for a reasonably reliable result one needs to integrate up to 
rather large values of $\alpha_2$.  We do this integral using the adaptive 
routine D01ASF for sine
and cosine transforms from the NAG Fortran library. 

We shall now turn to the question of the unitarity of the scattering amplitude
near the lowest threshold.  As mentioned in the Introduction, this is an 
interesting quantity to examine because one expects
non-perturbative effects to be particularly strong at thresholds.
Below the first inelastic threshold, unitarity implies that the total elastic
cross section is related to the forward scattering amplitude.  In the centre
of mass system
\begin{equation}
\sigma_{\rm el} ( s ) \> = \frac{1}{2 |{\bf p}| \sqrt{s}} \> 
{\rm Im} \> A(s,0)\;\;\; , 
\label{eq:optical}
\end{equation}
where ${\bf p}$ is the centre of mass three-momentum and the elastic 
cross section is given by
\begin{equation}
\sigma_{\rm el} ( s )  \> = \> \int d\Omega \> {1 \over 64 \pi^2 s} 
\> | A(s,t) |^2\;\;. 
\label{eq:elastic}
\end{equation}
Near threshold the elastic scattering cross section is, of course, dominated
by the real part of the amplitude because the imaginary part has to rise 
linearly with momentum from zero at threshold.
This means that $\sigma_{\rm el}$ is not particularly sensitive to
the exact position of the threshold.  For the imaginary part of the
forward scattering amplitude, however, the
threshold position is of supreme importance.  As was seen at zeroth order
in (III), a linear rise with 
$ |{\bf p}| $ as well as a correct threshold position was not
assured for an arbitrary trial action.  In particular, the large-$\sigma$ 
behaviour
of the retardation function entering this action was found to be crucial.

At first order one does not expect this to be quite as severe, as
at least agreement with one-loop perturbation theory (which has the correct
threshold behaviour) {\it is} assured for an arbitrary
trial action.  Nevertheless, the basic point remains the same: threshold
behaviour is governed by the large distance 
behaviour of the action and therefore we expect some sensitivity to the
large - $\sigma$ behaviour of the retardation function.  Hence
it is advisable to spend some time examining the dependence of the
imaginary part on the form of the profile function $A(E)$.

\subsection{${\cal I}m A(s,0)$ near  Threshold}
\label{sec:threshold rise}

Before doing this, it is actually very instructive to have a look at
${\cal I}m A(s,0)$ in 
perturbation theory.  To leading order in $ |{\bf p}| $, for each of the 
diagrams which includes a vertex correction, one obtains near the threshold
\begin{equation}
{\rm Im} \> A_V^{\rm pert.}(s,0) \> = \> -32 \alpha^2 \pi \>  
{M^4 \over (4 M^2 - m^2) (M+m) m^2}\> |{\bf p}| \> \> + \>
{\cal O} \left ( {\bf p}^2 \right )\;\;\;,
\end{equation}
the term involving the self energy of the intermediate propagator is
\begin{equation}
{\rm Im} \> A_P^{\rm pert.}(s,0) \> = \> 32 \alpha^2 \pi \> 
{M^4 \over (2 M + m)^2 (M+m) m^2}\> |{\bf p}|\> \> + \>
{\cal O} \left ( {\bf p}^2 \right )\;\;\;,
\end{equation}
and the box diagram yields
\begin{equation}
{\rm Im} \> A_B^{\rm pert.}(s,0) \> = \> 32 \alpha^2 \pi \>  
{M^4 \over (2 M - m)^2 (M+m) m^2}\> |{\bf p}|\> \> + \>
{\cal O} \left ( {\bf p}^2 \right )\;\;\;.
\end{equation}
Altogether, the leading contribution to the imaginary part is therefore
given by
\begin{equation}
{\rm Im} \> A^{\rm pert.}(s,0) \> = \> 128 \alpha^2 \pi \>  
{M^4 \over (4 M^2 - m^2)^2 (M+m)}\>  |{\bf p}|\> \> + \>
{\cal O} \left ( {\bf p}^2 \right )\;\;\;.
\label{eq:Im A pert}
\end{equation}
Even though each diagram separately is divergent as the meson mass $m$ goes
to zero, this divergence cancels in the complete amplitude.  
For the physical value of ${m \over M} \approx {1 \over 7}$ this means that
the individual diagrams are something like 50 times as large as the final
result.  This makes the task of calculating the imaginary part of the
variational expression for the forward scattering amplitude in 
Eq.~(\ref{eq:foramp}) even more formidable than it already is -- not only
does one have the numerical problems mentioned above, one also should
expect large cancelations between the various components of 
$ \> \Xi(s,t;\alpha_2) \> $.

At first sight it may appear that the lowest threshold will open up
as soon as $s$ is large enough so that one needs to work with the 
complex representation
in Eq.~(\ref{eq:foramp}) rather than the purely real representation
of Eq.~(\ref{eq:diramp}), i.e. $s > \tilde s(s)$.
Certainly, for $0 < s < \tilde s(s)$ the amplitude must be real, because
in this region  Eq.~(\ref{eq:diramp}) is valid.  It is useful to examine
the reason why  Eq.~(\ref{eq:foramp}), which is also valid in this
region of $s$, does not develop an imaginary part here.  
To this end we
shall define
\begin{equation}
{\cal D}(s) = \frac{1}{2} \left [ \> 1 - (1 - \lambda_2)^2 \> \right ] \> 
[ s - \tilde s(s) ] 
\label{def cal D}
\end{equation}
and
\begin{equation}
G(\alpha_2) =  \exp \left \{ \> {g^2 \over 8 \pi^2}\>   \Xi (s,t; i\alpha_2) 
 \> \right \} - 1\;\;\;.
\end{equation}
Hence the imaginary part of the forward scattering amplitude is
related to
\begin{equation}
\int_0^\infty d\alpha_2 \> \left [ \> \cos (\alpha_2 {\cal D})\> {\cal R}e 
\>G(\alpha_2) -
\sin (\alpha_2 {\cal D}) \>{\cal I}m \>G(\alpha_2) \> \right ] \;\;\;.
\label{eq:imsimp}
\end{equation}
Using the facts that $G(\alpha_2) = G^{\ast}(-\alpha_2)$, that 
$G(\alpha_2)$ has a cut along the positive imaginary axis and that
it vanishes sufficiently fast as $\alpha_2 \rightarrow \infty$, one
may easily prove, using dispersion methods, that
\begin{equation}
\int_0^\infty d\alpha_2 \> \cos (\alpha_2 {\cal D})\> {\cal R}e\> 
G(\alpha_2) = - \int_0^\infty d\alpha_2 \>\sin (\alpha_2 |{\cal D}|)\> 
{\cal I}m \>G(\alpha_2)\;\;\;.
\end{equation}
Hence the two contributions to the imaginary part of the scattering
amplitude in Eq.~(\ref{eq:imsimp}) cancel each other when
${\cal D} < 0$, i.e. $s < \tilde s(s)$, as required.  For $s > \tilde s(s)$
they remain identical in magnitude, but add.  This is a useful property
as it means that the amount of computer time required in the numerical 
integrations involved in the calculation 
of the imaginary part of Eq.~(\ref{eq:foramp}) is cut by a factor of 2.

Just because Eq.~(\ref{eq:foramp}) {\it may} develop an imaginary part
for $s$ above $\tilde s$ does not mean that it {\it will} do so.
In fact, one may use a similar argument to that used in the investigation
of the threshold position in (III) to show that the imaginary part
only starts somewhat above $\tilde s$.  Let us write $\Xi(s,0;\alpha_2)$
in Eq.~(\ref{eq:diramp}) as a Laplace transform:
\begin{equation}
{g^2 \over 8 \pi^2}  \> \Xi (s,0;\alpha_2) \equiv \int_0^\infty dE \>
{\tilde \rho}(E,s) \>e^{-E \alpha_2}\;\;\;.
\label{Laplace}
\end{equation}
Here ${\tilde \rho}(E,s)$ is essentially the counterpart of the weight function
$\rho(E)$ encountered in the zeroth order calculation in (III). It was found,
in fact, that $\rho(E)$ is only non-zero for $E$ larger than a critical
value $E_0$ and this will be also true for $\tilde{\rho}$, as will be shown
below. Expanding the exponential in Eq.~(\ref{eq:diramp}) with $s \rightarrow
s+i\epsilon$ and performing the $\alpha_2$ - integration then yields
\begin{equation}
{\cal I}m \> A_{2,2}^{\rm direct}(s,0)= \pi {\cal F}(s) 
\>\sum_{n=1}^{\infty} \> \frac{1}{n !} \> 
\int_{E_0}^{\infty} dE_1 ... dE_n\> {\tilde \rho}(E_1,s) ... 
{\tilde \rho}(E_n,s)
\>\delta \left (\sum_{i=1}^{n}E_i - {\cal D}(s) \right )
\label{eq:ima_expanded}
\end{equation}
Only when the $\delta$-function is fulfilled an imaginary part develops.
With this observation we proceed to  obtain an analytic expression for the 
leading behavior of the 
imaginary part of the amplitude near threshold.  Given the difficulties with
the numerical evaluation of Eq.~(\ref{eq:foramp}), in particular in view of
the large cancelations expected between $\Xi_V$, $\Xi_P$ and $\Xi_B$, the
possibility of having an analytic expression is
indeed fortunate.  
The key to obtaining this threshold behaviour is the observation that below
the first inelastic threshold  only the $n = 1$ term in 
Eq.~(\ref{eq:ima_expanded}) contributes.  This amounts to expanding the 
exponential in the expression for the amplitude Eq.~(\ref{eq:diramp}) 
and keeping only the first term; i.e.
\begin{equation}
{\cal I}m \> A_{2,2}^{\rm direct} (s,0)= {g^2 \over 8 \pi^2} \> {\cal F}(s)
\>{\cal I}m  \int_0^\infty d\alpha_2 \>
e^{\alpha_2 {\cal D}(s)}  \>
\Xi (s,0;\alpha_2)\hspace{1.5cm} E_0 < {\cal D}(s) < 2 E_0 \;\;.
\label{eq:imampth}
\end{equation}
Below the first inelastic threshold the imaginary part of the forward amplitude
is then simply given by
\be
{\cal I}m A_{2,2}^{\rm direct}(s,0)=  {\cal F} (s) \> 
\tilde{\rho} \left ( \> {\cal D}(s),s \> \right ) \> \Theta \left ({\cal D}(s) 
- E_0 \right )
\label{eq:ima_expanded2}
\ee
where we made use of the Laplace transform of $\Xi$ defined in
Eq.~(\ref{Laplace}).

In practice, the weight function $\tilde{\rho}(E,s)$ can be determined neither 
numerically (as inverse Laplace transforms are notoriously unstable) 
nor analytically because of the complicated $\alpha_2$ - dependence of 
the function $\Xi$. The only case where the inverse Laplace transformation 
can be performed is near threshold
which is known to be dominated by large values of $\alpha_2$. In this case we
can approximate the pseudotime by its asymptotic behaviour
\be
\mu^2(\alpha_2) \> \buildrel \alpha_2 \to \infty \over \longrightarrow \>
\frac{\alpha_2}{A(0)} + 4 \xi(0) + {\cal O} \left ( e^{-\alpha_2} \right )
\;\;\;.
\ee
This allows us to evaluate the corresponding asymptotic expressions for the 
$ \Xi$-functions. For example,  the propagator diagram's contribution
becomes
\bea
\Xi_P (\alpha_2) &\to& A(0) \int_0^{\infty} d\sigma \int_0^1 dx_1 \> 
\frac{\sigma}{\alpha_2 +
\sigma + 4 \xi(0)} \>  \exp \left [ - \left ( \frac{m^2}{2 A(0)} 
\frac{1-x_1}{x_1} + \frac{A(0)}{2}
\lambda_2^2 s x_1 \right ) ( \alpha_2 + \sigma) \right ] \nonumber \\
&& \hspace{4cm} \cdot \exp \left[ - 4 A(0) \xi(0) \left ( \frac{m^2}{2 A(0)} 
\frac{1-x_1}{x_1} - \frac{A(0)}{2}\lambda_2^2 \>s\> x_1 \right ) \right ] \>.
\eea
The above form is not a fully consistent expansion in inverse powers of
$1/\alpha_2$ since the exponential function will also contribute to such
terms. Nevertheless it is a convenient form to perform the inverse Laplace 
transform as we may use (Ref. \cite{TIT}, Eq. (5.5.9))
\be
{\cal L}^{-1} \left ( \> \frac{e^{-a \alpha_2}}{\alpha_2 + b} \> ; \alpha_2 
\> \right ) (E) \> = \> e^{-b (E-a)} \> \Theta(E-a)
\label{inverse Laplace}
\ee
to obtain
\be
\tilde{\rho}_P(E,s) \simeq \frac{g^2}{8\pi^2}\frac{A(0)}{E^2} \int_0^1 dx_1 \> 
\Theta \left (\> E - E_0(s,x_1) \>
\right )\> \exp \left [ \> - Q^2(s,x_1,E) \> \xi(0) \> \right ] \> .
\label{RP with x_1-int}
\ee
Here 
\be
E_0(s,x_1) \> = \> \frac{m^2}{2 A(0)}\frac{1-x_1}{x_1} + \frac{A(0)}{2}
\lambda_2^2 \>s \>x_1 
\label{def E0}
\ee
determines the minimal value of $E$ (see below) and
\be
Q^2(s,x_1,E) = 4 A(0) \left [ \> \frac{m^2}{2 A(0)}\frac{1-x_1}{x_1} - 
\frac{A(0)}{2} \lambda_2^2 s x_1 + E - E_0(s,x_1) \> \right ]
\label{def Q2}
\ee
is a kind of effective momentum transfer (one should remember from
(III) that $ \> 6 \> \xi(0) \> $ is the mean square radius of the dressed 
particle in lowest variational order ).

\noindent
Similarly we obtain for the asymptotic form of $\> \Xi\> $ in the box diagram
\bea
\Xi_B (\alpha_2) &\to& A(0) \int_0^{\infty} d\sigma \> \int_0^1 dx_1 \>  
\frac{\sigma}{\alpha_2 + \sigma + 4 \xi(0)} \> 
\exp \left[  \> - E_0(s,x_1) \alpha_2\> -  Q^2(s,x_1,E_0) \> \xi(0) \> \right ]
\nonumber \\
&&  \hspace{-0.2cm} \cdot \exp \left \{   
 - \left [ \frac{m^2}{2 A(0)}
\frac{1-x_1}{x_1} + \frac{A(0) x_1 \lambda_2 }{2} \left ( \lambda ( s + M^2 
- m^2 ) - \lambda_2 s \right ) \right ] \sigma  \> \right \} 
\eea
where again we have neglected terms of order $1/\alpha_2$ in the argument of 
the exponential function. Using the formula (\ref{inverse Laplace}) we get
\be
\tilde{\rho}_B(E,s) \simeq  \frac{g^2}{8\pi^2}\int_0^1 dx_1 \> \Theta \left ( 
\> E - E_0(s,x_1) \> \right ) \> 
\frac{A(0)}{E_1^2(s,x_1,E)}
\> \exp \left [ \> - Q^2(s,x_1,E) \> \xi(0)  \> \right ] \> .
\label{RB with x_1-int}
\ee
\noindent
where
\be
E_1(s,x_1,E) \> = \> \frac{m^2}{2 A(0)}\frac{1-x_1}{x_1} + \frac{A(0) x_1 
\lambda_2}{2}
\left ( \lambda ( s + M^2 - m^2) - \lambda_2 s \right ) + E - E_0(s,x_1) \> .
\label{def E1}
\ee
\noindent
Finally the $\Xi$-function in the vertex diagram has the asymptotic limit
\bea
\Xi_V (\alpha_2) &\to& - A(0) \int_0^{\infty} d\sigma \>  \int_0^1 dx_1 
dx_2 \>  \frac{\sigma}{\alpha_2 + \sigma + 4 \xi(0)} \> 
\exp \left[   \> - E_0(s,x_1) \alpha_2  - 
 Q^2(s,x_1,E_0) \> \xi(0)\> \right]
\nonumber \\
&&   \cdot \exp \left \{
- \left [ \frac{m^2}{2 A(0)}
\frac{1-x_1}{x_1} + \frac{A(0) x_1 \lambda_2 }{2} \left ( \lambda x_2 ( s + 
M^2 - m^2 ) +
(1-2x_2)\lambda_2 s \right ) \right ] \sigma  \> \right \} \> .
\eea
Consequently the corresponding weight function for the {\it two} identical
vertex contributions becomes
\bea
\tilde{\rho}_V(E,s) &\simeq& - 2\frac{g^2}{8\pi^2} A(0)\int_0^1 dx_1 \> dx_2 \> 
\Theta \left ( \> E - E_0(s,x_1) \> \right ) \>
\> \exp \left [ \> - Q^2(s,x_1,E) \> \xi(0) \> \right ] \nonumber \\
&& \hspace{0.5cm} \cdot \left [ \> \frac{m^2}{2 A(0)}
\frac{1-x_1}{x_1} + \frac{A(0) x_1 \lambda_2 }{2} \left ( \lambda x_2 (s + 
M^2 - m^2) + (1-2 x_2)\lambda_2 s ) \right )  \> \right ]^{-1} .
\eea
The $x_2$-integral is the same as the one for combining denominators 
by the Feynman parameterization 
\be
\int_0^1 dx_2 \> \frac{1}{[ \> a + b x_2 \> ]^2 } \> = \> \frac{1}{a b}
\ee
and thus we obtain
\be
\tilde{\rho}_V(E,s) \simeq - 2\frac{g^2}{8\pi^2} A(0)\int_0^1 dx_1 \> \Theta 
\left ( \> E - E_0(s,x_1) \> \right ) 
\> \frac{1}{E E_1(s,x_1,E) }
\> \exp \left [ \> - Q^2(s,x_1,E) \xi(0) \> \right ] \> .
\label{RV with x_1-int}
\ee
Combining all terms gives the simple formula
\be
\tilde{\rho}(E,s) \simeq \frac{g^2}{8\pi^2}A(0)\int_0^1 dx_1 \> \Theta \left ( 
\> E - E_0(s,x_1) \> \right ) \> 
\left ( \frac{1}{E} - \frac{1}{E_1(s,x_1,E)} \right )^2
\> \exp \left [ \> - Q^2(s,x_1,E) \xi(0) \> \right ] 
\label{R with x_1-int}
\ee
which is evidently positive.   This equation, together with 
Eq.~(\ref{eq:ima_expanded2}), directly determines
the imaginary part of the forward scattering amplitude.  Hence one obtains
 a positive total cross section, as required.  In the following,
we shall first examine for which $s$ this imaginary part first appears and
subsequently shall derive an {\it analytical} expression for the total
cross section at threshold.

\subsection{Threshold Position}
\label{sec:threshold}

Let us now determine the threshold position. Eq. (\ref{R with x_1-int}) 
should hold in the vicinity of the elastic threshold.
Approaching it, the range of the $x_1$-integration shrinks due to the step 
function and is restricted to the interval 
$ \> x_1^{(-)} \> < x_1 \> < x_1^{(+)} \> $ where
\be
x_1^{(\pm)} \> = \> \frac{1}{A(0) \> \lambda_2^2 \>  s} \> \left [ \> E +
\frac{m^2}{2 A(0)}
\> \pm \> \sqrt{ \left (E + \frac{m^2}{2 A(0)} \right )^2 - m^2\> 
\lambda_2^2 \> s }
\> \> \right ] \> .
\label{u12}
\ee
Exactly at threshold $ x_1^{(+)} $ and $x_1^{(-)} $ coincide and only the 
minimum point
\be
\bar x_1 = \frac{m}{A(0) \lambda_2 \sqrt{s}}
\label{u0}
\ee
is left over. Eq. (\ref{u0}) determines the lowest threshold position to be
\be
E_{\rm thr.} = {\cal D}(s_{\rm thr.}) = E_0(\bar x_1,s_{\rm thr.}) 
\ee
or 
\begin{equation}
E_0 = \lambda m \sqrt s - {m^2 \over 2 A(0)}\;\;\;.
\label{eq:ezero}
\ee
which is  the statement that  that ${\tilde \rho}(E,s)$ is only
non-zero for $E$ larger than $E_0$.         
Inserting Eq.~(\ref{eq:ezero}) into Eq.~(\ref{eq:ima_expanded}), one finds  
the $n^{\rm th}$  threshold condition 
\begin{equation}
  n \left (2 \> \lambda_2 \>  m \sqrt s_{\rm thr.} - {m^2 \over  A(0)}\right ) 
= \left [ \> 1 - (1 - \lambda_2)^2 \> \right ] \> 
\left [ \> s_{\rm thr.}- \tilde s(s_{\rm thr.}) \> \right ] \;\;\;.
\label{eq:constraint}
\end{equation}

The parameter $\lambda_2$ is undetermined up to now.  Certainly, one might 
try to fix it
via the variational principle, which, however, in practice would seem to be
rather difficult.  An alternative, which is what we shall do here, is to
fix it by {\it demanding} that the lowest threshold (i.e. $n$ = 1)
coincides exactly with $\sqrt s_{\rm thr.}$ = $M+m$. \footnote{Another 
alternative, again something which we do not pursue here, is to
allow $\lambda_2$ to become a function of $s$ and hereby enforce all
threshold positions to be precisely where they should be.}

\noindent
In order to solve for $\lambda_2$, we rewrite Eq.~(\ref{eq:constraint})
as
\be
\lambda_2^{(\pm)} \> = \> \frac{1}{M +  m} \> \left [ \> M 
\> {\pm} \> \sqrt { M^2 (1-\lambda)^2 + \frac{m^2}{A(0)} -
\frac{\alpha M^2}{\pi}\Xi_p^1(s=(M+m)^2)} \> \right ] \>.
\label{eq:threshold_exact}
\ee
As indicated, in general there are at least two solutions for $\lambda_2$.
In fact, because $ \> \Xi_p^1\> $ also depends on $\lambda_2$,
this equation is only implicit and actually may result in
more than two solutions which have to be determined numerically. 
We have solved the implicit equation by iteration and found
that below a critical coupling of about $\alpha = 0.4$
each branch $\lambda_2^{(+)}$ and $\lambda_2^{(-)}$ yields exactly one
solution. Above this coupling the solution for $\lambda_2^{(+)}$
ceases to exist while at the same time  the branch $\lambda_2^{(-)}$
develops another solution.  As the coupling is increased further, more
solutions for $\lambda_2^{(-)}$ come into existence.

Clearly, one needs to decide which one of these solutions corresponds
to reality.  For small couplings the decision seems relatively
easy -- the
perturbative limit (in which $\lambda  = A(0) = 1$ and $\tilde 
s = M^2$) of Eq.~(\ref{eq:threshold_exact}) is
\begin{equation}
{\lambda_2}^{(\pm)} \> = \> {M \pm m  \over M + m} \> + \>
{\cal O} \left (\alpha \right )\;\;\;.
\label{eq:small_alpha}
\end{equation}
As mentioned before, we require that in the limit of
zero coupling the trial action becomes equal to the true action.  
Clearly only $\lambda_2^{(+)}$ in the above equation satisfies this
requirement, so we shall use this solution up to the critical coupling
of $\approx 0.4$.  Beyond this coupling we shall contend ourselves
with using that value of $\lambda_2^{(-)}$ which matches smoothly
onto $\lambda_2^{(+)}$ (only one of the solutions for $\lambda_2^{(-)}$
has this property).  

It is not entirely clear whether this is a satisfactory choice.  In
particular, one is naturally led to speculate whether this critical
coupling actually has a physical meaning -- perhaps the intermediate
off-shell propagator feels the instability of the Wick-Cutkosky model
already at smaller coupling constant than the on-shell propagator.
\footnote{We remind the reader that the
variational equations for the on-shell propagator cease to have real solutions
for couplings greater than about $\alpha = 0.8$, signaling the
instability associated with this model.}  On the other hand, the
division of solutions into $\lambda_2^{(+)}$ and $\lambda_2^{(-)}$ in
Eq.~(\ref{eq:threshold_exact}) is somewhat artificial as, after all,
this equation is only an implicit one for $\lambda_2$.  Hence it is
also possible that the critical coupling is just a mathematical
artifact.

\subsection{Total cross section}

We now turn to the total cross section.  Very close to threshold we can write
\bea
\tilde{\rho}(E,s)  &\simeq& \frac{g^2}{8\pi^2}
A(0) \left ( x_1^{(+)} - x_1^{(-)} \right ) 
\left (
\frac{1}{E} - \frac{1}{E_1(s_{\rm thr.},\bar x_1,E_0)} \right )^2
\> \exp \left [ \> - Q^2(s_{\rm thr.},\bar x_1,E_0) \xi(0) \> \right ] 
\nonumber \\
&\simeq& \frac{g^2}{8\pi^2} \frac{2}{\lambda_2^2 (M+m)^2} \sqrt{ \left [ E - 
E_0(s_{\rm thr.},\bar x_1) \right ] 
\>  2m \lambda_2 (M+m) } \> \> e^{2 m^2 \xi(0)} 
\nonumber \\
&& \hspace{3cm} \cdot \left (
\frac{1}{E_0(s_{\rm thr.},\bar x_1)} - \frac{1}{E_1(s_{\rm thr.},\bar x_1,E_0)} 
\right )^2 
\label{R without u-int}
\eea
where
\bea
E_0(s_{\rm thr.},\bar x_1) \> = \> m \lambda_2 (M + m) - \frac{m^2}{2 A_0} 
\> \equiv \> E_0\\
E_1(s_{\rm thr.},\bar x_1,E_0) \> = \> m \lambda M - \frac{m^2}{2 A_0} 
\> \equiv \> E_1 \> .
\eea
The square root $ \sqrt{E - E_0} $ gives rise to a {\it linear}
behaviour of the imaginary part of the amplitude with the 
three-momentum ${\bf p}$ in the center-of-mass system
as can be seen by expanding
\be
s = \left ( \> \sqrt{M^2 + {\bf p}^2} + \sqrt{m^2 + {\bf p}^2} \> \right )^2 
\> .
\ee
In contrast to the zeroth order variational approximation (III)
where this behaviour was crucially dependent on the specific form
of the retardation function at large values of $\sigma$, in first order
this now holds for {\it all} parameterizations. This demonstrates the
increased insensitivity of the first-order result  to the
variational {\it ansatz}. 
At threshold one obtains 
\begin{equation}
{\cal I}m \> A(s_{\rm thr.},0) = {\alpha \>{\cal F}(s_{\rm thr.}) \>M^2 \>
e^{2 m^2 \xi(0)} \over
\sqrt{ (M+m) M } \> \lambda_2 m^2}
\>{\cal Z} \> { \left [ \lambda_2 m + (\lambda_2 - \lambda) M \right ]^2
\over \left [ \lambda_2 (M+m) - {m \over 2 A(0)}\right ]^2 \> 
\left ( \lambda M - {m \over 2 A(0)}\right )^2 }\> p\;\;,
\label{eq:imampth6}
\end{equation}
where the derivative in
\begin{equation}
{\cal Z}\> = \>\sqrt{2 - \lambda_2 - {m \over m+M} - 
{\alpha M^2 \over \pi \lambda_2} \> 
{d \over ds}\Xi_P^1(s) \> }
\end{equation} 
is to be evaluated at threshold. By the optical theorem the total
cross section at threshold is obtained from $ {\cal I}m \> A(s,0)$ using
Eq.~(\ref{eq:optical}). It may readily be verified that this expression 
reduces to Eq.~(\ref{eq:Im A pert}) in the perturbative limit.

\section{Numerical Results}
\label{sec:numerics}
\setcounter{equation}{0}

In Table~1 we collect the variational parameters determined from the
variational equations at the pole of the propagator as well as the
values for $\lambda_2$  fixed at the lowest threshold. 
As usual we have taken $ M = 939 $ MeV and $ m = 140 $ MeV for the 
mass of the nucleon and meson, repectively. It should be noted that the 
parameters of the present `extended' parameterizations are slighthly 
different from those given in Ref. (III) because the variation now is not
constrained to give the correct (zeroth order) thresholds. Consequently
the minimum values of the variational functional are always lower than 
previously and are nearly identical with the full variational calculation
(assuming no specific functional form for the retardation function), except
at the largest coupling constant.

The total cross section at threshold is evaluated using the optical
theorem and the analytic result of Eq.~(\ref{eq:imampth6}). The elastic
cross section at threshold is determined by the real part of the
scattering amplitude which is evaluated numerically. The `unitarity
ratio' $ \> \sigma_{\rm el}/\sigma_{\rm tot} \> $ is shown in 
Fig.~\ref{fig: sigratio} for 
the  allowed range of couplings. For comparison we include the results
obtained in first order perturbation theory.  As can be seen the
variational result remains close to the unity whereas, as is well
known,
perturbation theory strongly violates it for large couplings.

\begin{center}
\begin{tabular}{|l|c|c|c|c|c|c|c|c|} \hline
                   &                &                &                &  
                   &                &                &                & \\
                   & $\alpha = 0.1$ & $\alpha = 0.2$ & $\alpha = 0.3$ & 
                     $\alpha = 0.4$ & $\alpha = 0.5$ & $\alpha = 0.6$ & 
                     $\alpha = 0.7$ & $\alpha = 0.8$ \\
                   &                &                &              &
                   &                &                &              & \\ \hline
                   &                &                &              &  
                   &                &                &              & \\
 $y_1$             & 1.01142        & 1.01950        &  1.02899     & 1.04023
                   & 1.05405        & 1.07171        &  1.09755     & 1.15359 \\
 $y_2$             & 2.0702         & 2.3960         &  2.6302      & 2.8824
                   & 3.0850         & 3.1742         &  3.2194      & 3.1266 \\
 $\sqrt{w_1}$ ~[MeV] & 636.09       & 612.10         &  587.31      & 560.04
                   & 530.05         & 496.02         &  453.33      & 381.08 \\
 $\sqrt{w_2}$ ~[MeV] & 376.68       & 370.04         &  358.92      & 345.96
                   & 329.66         & 308.04         &  280.66      & 233.72 \\
 $\lambda$         & 0.97297        & 0.94390        &  0.91223     & 0.87718 
                   & 0.83739        & 0.79032        &  0.72977     & 0.62361 \\
 $\lambda_2$       & 0.97690        & 0.95215        &  0.92534     & 0.89586 
                   & 0.86273        & 0.82414        &  0.77587     & 0.69838 \\
                   &                &                &              &
                   &                &                &              & \\ \hline 
                   &                &                &              &
                   &                &                &              & \\
 $A(0)$            & 1.01508        & 1.03213        &  1.05177     & 1.07489 
                   & 1.10312        & 1.13951        &  1.19203     & 1.30393 \\
 $Z$               & 0.96087        & 0.91918        &  0.87429     & 0.82524 
                   & 0.77045        & 0.70693        &  0.62748     & 0.49509 \\
 $\left <r^2\right >_0^{1/2}$ [fm] & 0.0182 & 0.0258 &  0.0313      & 0.0352
                   & 0.0368         & 0.0327         &  -0.0164     & -0.0886 \\
 $\left <r^2\right >_1^{1/2}$ [fm] & 0.0481 & 0.0697 &  0.0879      & 0.1049  
                   & 0.1220         & 0.1404         &  0.1623      &0.1983  \\
                   &                &                &              & 
                   &                &                &              & \\ \hline
\end{tabular}

\end{center}

\vspace{0.5cm}
\noindent
Table 1 :
The variational parameters for the` extended' retardation
function writing  $ \> C_1 = g^2 y_1/(32\pi^2) \> $ and $ \> C_2 = y_2
m(m/M)^{3/2}\sqrt{\pi/2} \> $ and some derived quantities.
A negative zeroth order rms-radius indicates
that $\left <r^2\right >_0 \> \> < \> \> 0 $.

\vspace{0.8cm}

\begin{figure}[hbtp]
\begin{center}
\mbox{\epsfxsize=8cm\epsfysize=8cm\epsffile{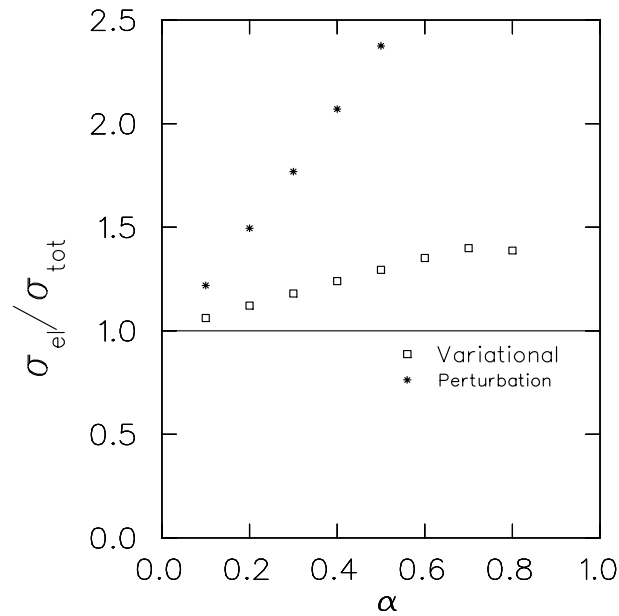}}
\end{center}
\caption{The unitarity ratio $\sigma_{\rm el}/\sigma_{\rm tot}$ at
threshold as a function of the coupling $\alpha=g^2/(4\pi M^2)$ .}
\label{fig: sigratio}
\end{figure}

We also investigated the momentum dependence of the total cross section
above the elastic, but below the first inelastic, threshold. 
For very small momenta one can
apply Eq.~(\ref{R with x_1-int}) to obtain  an analytic expression for 
the imaginary part of the forward
scattering amplitude as a function of momentum  as
was done at threshold.
 In Fig.~\ref{fig: total} we show the results for
$\alpha=0.8$. The results for the three lowest momenta are calculated
analytically and they join nicely to the results obtained numerically 
at larger momenta.  In contrast to what was observed for the vertex function
in Ref. (IV), here one finds a  smooth behaviour of the total cross section
above threshold as a function of the momentum.

\begin{figure}[hbtp]
\begin{center}
\mbox{\epsfysize=8cm\epsffile{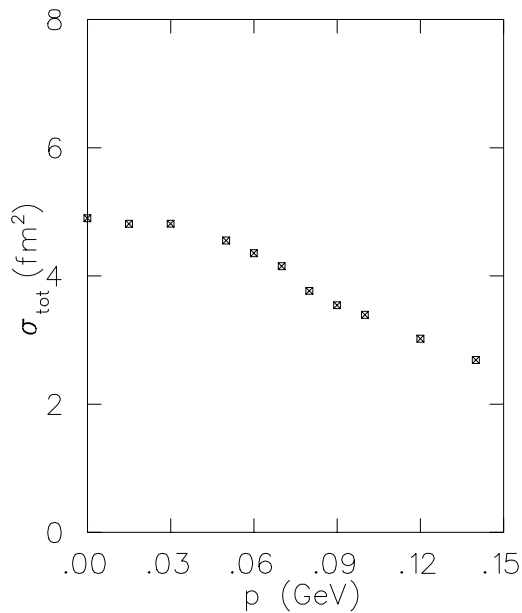}}
\end{center}
\caption{The total cross section as a function of the center-of-mass 
momentum $ |{\bf p}| $ at $\alpha=0.8$ .}
\label{fig: total}
\end{figure}

\section{Conclusions and Outlook}
\label{sec:conclusion}

In this work Feynman's variational method originally applied to the
polaron problem, a non-relativistic field theory, was successfully
generalized to study particle scattering in a relativistic setting.  
A central feature of the approach is that the variational principle is
applied in the particle (or worldline) representation of field theory, 
making it rather similar to the non-relativistic theory.  This is
particularly so for the scalar theory under discussion in this paper
-- it is fair to characterise it as essentially Feynman's polaron in
$4+1$ rather than $3+1$ dimensions (with, in this case, phonons of
non-zero mass).

The method amounts to a well-defined non-perturbative expansion of the
Green functions of the theory.  The leading term, which is the one
calculated in this paper, obeys a {\it variational principle}.  Higher order
terms could be calculated as they have been for the
polaron~\cite{LuRo}, if the need arose. Hence, roughly speaking, the
``expansion parameter'' is a dynamically determined quantity which
adjusts itself in order to minimize the corrections due to these higher
order terms. It is, of course, crucial, that the starting point - the trial
action - contains as much of the physics of the problem as possible.
Although practically we are limited to quadratic (but {\it nonlocal}) trial
actions there seems to be sufficient flexibility in the {\it ansatz} to 
achieve this goal. In the present work we even have generalized the previous 
trial action for better off-shell propagation without destroying the 
advantageous features of our method: the variational equations remain the
same as previously, are independent of the number of external mesons
and can be easily solved in Euclidean space.

In any expansion scheme such as this it is desirable that the lowest
order term(s) are not only numerically reliable, but that the
expansion also exhibits some of the analytic features of the full
theory.  A trivial illustration of this statement is relativistic
invariance: both perturbation theory, for example, as well as the
method outlined here are written in a manifestly invariant form. Also,
crossing symmetry is maintained in both approaches.  Another example,
one that perturbation theory does not share with the present approach,
is that the lowest order term already exhibits all the cuts and poles
of the exact Green function.

A property of the theory which is connected with this, and which is 
maintained (at finite order) neither in the variational calculation nor in
perturbation theory is the unitarity of the theory.  In particular,
the imaginary part of the forward scattering amplitude is connected to 
the total cross section via the optical theorem.  Perturbation theory violates
this very badly, something which is of course most obvious at tree level --
here the imaginary part of the forward scattering cross section is identically
zero because of the lack of analytic structure mentioned above.
Because the variational calculation does not suffer from this deficiency
one might suspect that unitarity is partially restored.  The numerical
results presented in this paper confirm this.

Although this is an important result in itself, restoration of unitarity was
of course not the prime motivation for the work presented here.  Rather,
we were interested in the intrinsically non-perturbative nature of the
variational method.  For this reason we concentrated our attention on a
kinematic region where non-perturbative effects are usually most important,
namely near a threshold.  We chose the lowest theshold because this did not
require a calculation of higher-point Green functions.

The threshold region is important from another perspective as well.
This region is sensitive to the infrared behaviour of the theory,
while the fact that the theory is a divergent field theory means that
ultraviolet physics plays a dominant role as well.  The variational
method is able to deal sensibly with both these regions simultaneously
through the use of the profile function $A(E)$ in the trial action,
which is complementary to the retardation function
$f(\sigma)$. Specifically, the ultraviolet (infrared) behaviour is
simulated by the small- (large-) $\sigma$ behaviour of this function
which corresponds to the large- (small-) $E$ behaviour of $A(E)$.

A difficulty with investigating the behaviour of the Green function near
threshold is that although the variational approximation to the forward
scattering amplitude does indeed contain cuts which one can identify with
one, two etc. meson production in the intermediate state, the thresholds
for these cuts are in general only approximately equal to their physical values.
The solution of this problem which we chose illustrates another important
feature of the variational method:  although some properties of the
exact Green function are reproduced for arbitrary trial actions, others
may be enforced by restricting the form of the trial action.  In particular,
in the present work the new parameter $\lambda_2$, associated with the
internal nucleon line, was fixed so that the lowest threshold was
identical to the physical one.  

This is certainly a legitimate procedure, although it  would be more
satisfying (and one would expect the numerical results to improve) if the 
variational principle itself would provide the correct thresholds. This
would mean, in the context of our trial action, treating $\lambda_2$ 
 as a variational parameter with  the daunting task of solving
the variational equations for this parameter.  We have chosen not to go
into this direction but rather
concentrate our efforts on a generalization of the present approach 
to a more realistic field theory, namely Quantum Electrodynamics. 

\vspace{1cm}
\noindent
{\bf Acknowledgement:} We thank Nadia Fettes for a careful reading of the 
manuscript.

\end{document}